\documentclass[preprint]{aastex631}
\usepackage{amsmath}

\hypersetup{linkcolor=blue,citecolor=blue,filecolor=cyan,urlcolor=magenta}

\newcommand{\um}{$\mu$m}
\newcommand{\lv}{\left\vert}
\newcommand{\rv}{\right\vert}

\shorttitle{Mathematical Model for X-ray Telescope}
\shortauthors{Jiang et al.}

\graphicspath{{./}{figures/}}
\DeclareUnicodeCharacter{2061}{}

\begin{document}

\title{Construction and Validation of a Geometry-based Mathematical Model for Modulating Imaging X-ray Space Telescope}

\author{ Xian-Kai Jiang }
\affiliation{Purple Mountain Observatory, Chinese Academy of Sciences, 10 Yuanhua Road, Qixia District, Nanjing 210023, China}
\affiliation{School of Astronomy and Space Sciences, University of Science and Technology of China, Hefei 230026, China}

\author{ Jian Wu }
\affiliation{Purple Mountain Observatory, Chinese Academy of Sciences, 10 Yuanhua Road, Qixia District, Nanjing 210023, China}
\affiliation{School of Astronomy and Space Sciences, University of Science and Technology of China, Hefei 230026, China}

\author{ Deng-Yi Chen }
\affiliation{Purple Mountain Observatory, Chinese Academy of Sciences, 10 Yuanhua Road, Qixia District, Nanjing 210023, China}
\affiliation{School of Astronomy and Space Sciences, University of Science and Technology of China, Hefei 230026, China}

\author{ Yi-Ming Hu }
\affiliation{Purple Mountain Observatory, Chinese Academy of Sciences, 10 Yuanhua Road, Qixia District, Nanjing 210023, China}
\affiliation{School of Astronomy and Space Sciences, University of Science and Technology of China, Hefei 230026, China}

\author{ Hao-Xiang Wang }
\affiliation{Purple Mountain Observatory, Chinese Academy of Sciences, 10 Yuanhua Road, Qixia District, Nanjing 210023, China}
\affiliation{School of Astronomy and Space Sciences, University of Science and Technology of China, Hefei 230026, China}

\author{ Wei Liu }
\affiliation{Purple Mountain Observatory, Chinese Academy of Sciences, 10 Yuanhua Road, Qixia District, Nanjing 210023, China}
\affiliation{School of Astronomy and Space Sciences, University of Science and Technology of China, Hefei 230026, China}

\author{ Zhe Zhang }
\affiliation{Purple Mountain Observatory, Chinese Academy of Sciences, 10 Yuanhua Road, Qixia District, Nanjing 210023, China}
\affiliation{School of Astronomy and Space Sciences, University of Science and Technology of China, Hefei 230026, China}
\email{zhangzhe@pmo.ac.cn}

\begin{abstract}

Quantitative and analytical analysis of modulation process of the collimator is a great challenge, 
and is also of great value to the design and development of Fourier transform imaging telescopes.
The Hard X-ray Imager (HXI), as one of the three payloads onboard the Advanced Space-based Solar Observatory(ASO-S) mission,
adopts modulating Fourier-Transformation imaging technique and will be used to 
explore mechanism of energy release and transmission in solar flare activities.
In this paper, a mathematical model is developed to analyze the modulation function under a simplified condition first. 
Then its behavior under six degrees of freedom is calculated after adding the rotation matrix and translation change to the model.
In addition, unparalleled light and extended sources also are considered 
so that our model can be used to analyze the X-ray beam experiment. 
Finally, applied to the practical HXI conditions, the model has been confirmed 
not only by Geant4 simulations but also by some verification experiments.
The model has demonstrated powerful performance and advantages in instrument design and evaluation. 
Furthermore, it will help to improve the image reconstruction process after the launch of ASO-S.

\end{abstract}

\keywords{X-ray telescope --- the collimator --- Fourier-transform --- Bessel function --- Geant4 simulation --- X-ray beam experiment}

\section{Introduction} \label{sec:intro}

The imaging observation in the hard X-ray band is of great significance for understanding 
the mechanism of energy release and transmission in solar flare activities. 
The Hard X-ray Imager(HXI) is a new hard X-ray telescope that applies the Fourier-transform (FT) imaging technique.
As one of the indirect imaging instruments, it images the sun by reconstructing from its pattern and measured visibility\citep{su2019simulations}.
So, an accurate pattern is the basis for high-quality imaging. 
However, the current methods of calculating pattern are either approximate or numerical, 
which affects the analyticity and efficiency of the imaging algorithm to some extent.
In order to obtain a more accurate pattern, a mathematical model based on the geometric structure of HXI is developed in this paper.
And the necessity of the new method can be demonstrated by an in-depth analysis of the imaging technique.

Direct imaging and indirect imaging are the two major techniques to observe the sun in hard X-ray band.
But in terms of a hard X-ray observation instrument, direct imaging as that in the visible light band 
still faces many technological difficulties, although it has been developed for many years \citep{mi2019stacked}.
In contrast, Fourier-transform imaging, as a kind of typical indirect modulation imaging technique, 
has been applied to solar X-ray imaging observation and achieved fruitful results for several decades. 
The hard X-ray telescope(HXT) onboard the Yohkoh mission \citep{sakao1994characteristics}, 
the Reuven Ramaty High-Energy Solar Spectroscopic Imager(RHESSI) \citep{lin2003reuven},
and the Spectrometer/Telescope for Imaging X-rays(STIX) onboard the Solar Orbiter mission \citep{krucker2020spectrometer}
are three of the most well-known solar X-ray imaging instruments using such FT imaging technique. 
And the new solar hard X-ray imaging instrument HXI also applies the FT imaging technique.

To study the imaging process of HXI, it is first necessary to state the imaging principle of FT type telescopes.
This kind of instrument contains several groups of bi-grid sub-collimators with different pitches, position angles, and phase angles.
The differences of each sub-collimator can reflect the spatial information of the sources, which are called ``modulation modes".
Each sub-collimator modulates the photon flux physically, 
and the detector right behind it records those photons that pass through the sub-collimator as a result of the modulation. 
By subtracting the particle background, results containing only the original information component can be obtained.
The original image can be rebuilt by analyzing the modulation modes with their results.

The most popular way to achieve image reconstruction is to go through its pattern and visibility.
The pattern is formed by the angle-transmittance response, or called modulation function of the corresponding sub-collimator,
and the pattern is actually the final modulation function of the detector.
Meanwhile, the visibility is formed by the counting of a pair of sine and cosine sub-collimator with the same modulation mode,
each representing a two-dimensional Fourier component of the source.
After getting the pattern and visibility, further methods like Clean, Pixons, and Forward-Fitting, 
can be used to form them into reconstructed image \citep{hurford2003rhessi}. 
Although the mathematical details of those methods are totally different, they all start with the analysis of the pattern.
It is worth noting that the effect on the image quality caused by observation errors during the modulation and reconstruction process 
is independent of the effect caused by using an inaccurate pattern, and the inaccurate pattern amplifies errors from observation.
The modulation function serves as the basis of the entire reconstruction algorithm, 
a more precise pattern can lead to a better-reconstructed image.
Therefore, it is beneficial to obtain knowledge about modulation functions as much as possible.

The modulation function of a sub-collimator presents a simple triangle-wave form for the ideal instrument condition. 
However, the modulation function takes on a more complicated shape due to practical manufacturing techniques and environmental effects. 
Prince and Hurford mentioned a uniform way to analyze sub-collimators 
by convolving the two layers of the grid in their review \citep{prince1988gamma}. 
They also pointed out that in many cases, despite being a triangle wave, 
the modulation function can be considered as a cosine function. 
This was also used in the HXT data progress \citep{kosugi1991hard}.
Because the widths of the slits and slats of grids on RHESSI are unequal, the standard triangle wave could not be used. 
To improve the accuracy of the reconstruction, a Fourier cosine expansion was adapted as the modulation function \citep{hurford2003rhessi}. 
And this method of Fourier cosine expansion was also applied in STIX to form its pattern \citep{benz2012spectrometer}.

However, the precision of these methods is limited for a detailed analysis of actual engineering errors, 
such as machining tolerance, deformation of the instrument et al.
So, numerical simulation has been used as a complement to it, 
and the modulation function in special cases can be solved by performing Monte Carlo simulations.
But anyway, this would take more time to perform the simulation works,
especially under the condition that some parameters need to be changed within a certain range.
Although the numerical method is more accurate, it is less resolvable and efficient.
Consequently, since HXI has been implemented, there is an urgent need to develop a more detailed mathematical model of the collimator.
During the assembly and launch of the satellite, the collimator might undergo some slight deformation. 
These deformations will have a considerable impact on the imaging reconstruction process, 
so each sub-collimator's modulation function needs to be modified according to its deformation.
Traditional methods of computing modulation functions are either inaccurate or unanalytical,
but our accurate mathematical model can do better in terms of speed, accuracy, and analyticity,
and will certainly facilitate the evaluation of HXI's test and data process.

The new instrument HXI, as one of the payloads onboard 
the Advanced Space-based Solar Observatory (ASO-S), is an FT-type imaging telescope used to investigate 
the acceleration and transmission of electrons in the solar atmosphere 
during eruptions \citep{zhang2019hard,su2019simulations}. 
ASO-S is China's first comprehensive solar mission, launched on the ninth of October, 2022.
Aiming for the 25th solar maximum, it focuses on three major fields of solar activity: 
the photosphere magnetic field, coronal mass ejections (CMEs), and solar flares \citep{gan2019advanced}. 
This satellite has been sent to a 720-km Sun-synchronous orbit and will have a lifetime of at least four years.
HXI employs the spatial modulation technique that is similar to Yohkoh/HXT.
Table \ref{tab:mission} shows the configuration and characteristics of 
Yohkoh/HXT, RHESSI and SO/STIX in comparison to HXI. 

The rest of this paper is structured as follows: 
In Section \ref{sec:calculation}, a new mathematical model for the transmission function of a collimator 
which is based on the geometric relationship is introduced first step by step, 
and then put into a variety of complex conditions to see how it works.
Next, a brief introduction to the structure of HXI is given in Section \ref{sec:testment},
followed by the simulation and experimental works done with the practical HXI 
collimator conditions in order to confirm the mathematical model. 
Then it comes to a discussion about how this model helps the data process in Section \ref{subsec:discussion}.
Finally, Section \ref{sec:summary} gives a summary and a brief outlook on this work. 

\begin{center}
  \begin{table}[ht]
      \centering
      \caption{Comparisons of several missions with HXI}
      \footnotesize
      \label{tab:mission}
      \begin{tabular}{l c c c c}
          \hline\hline
          & YOHKOH/HXT & RHESSI & Solar Orbiter/STIX & ASO-S/HXI \\
          \hline
          Launch time & 1991 & 2002 & 2020 & 2022 \\
          Imaging method & SMC & RMC & SMC & SMC \\
          Quantity of sub-coll. & 64 & 9 & 32 & 91 \\
          Pitch of grids & finest 105 \um & 34 \um $\sim$ 2.75 mm & 38 \um $\sim$ 1 mm & 36 \um $\sim$ 1224 \um \\
          Spatial resolution & 5$\arcsec$ & 2.3$\arcsec\,$@100 keV & 7$\arcsec$ & 3$\arcsec$ \\ 
          Detector & NaI(Tl) & Ge & CdTe & LaBr$_3$ \\
          Energy range & 20 keV$\sim$ 100 keV & 3 keV$\sim$ 17 MeV & 4 keV$\sim$ 150 keV & 30 keV$\sim$ 200 keV \\
          Temporal resolution & 0.5s & 2s & up to 0.5s & up to 0.125s \\
          \hline
          \multicolumn{5}{l}{sub-coll: sub-collimator, SMC: space modulation collimator, RMC: rotate modulation collimator.} \\
      \end{tabular}
  \end{table}
\end{center}

\section{Calculation of Modulation Functions} \label{sec:calculation}

In this section, we start to build a detailed model for sub-collimators.
Firstly, a simple case with only a sub-collimator in perfect condition is discussed.
Next, the six degrees of freedom for a rigid body are calculated on this basis, 
and then the simplification of this model will be reported under some particular conditions. 
Finally, extended source and unparalleled light are considered to match the X-ray beam test.
This section contains a large number of symbolic operations, to make it clearer, 
a table explaining the various symbols can be found in Appendix \ref{sec:symbol}.

\subsection{Basic Calculation} \label{subsec:basic}

\begin{figure}[ht!]
  \centering
  \includegraphics[width=0.9\textwidth]{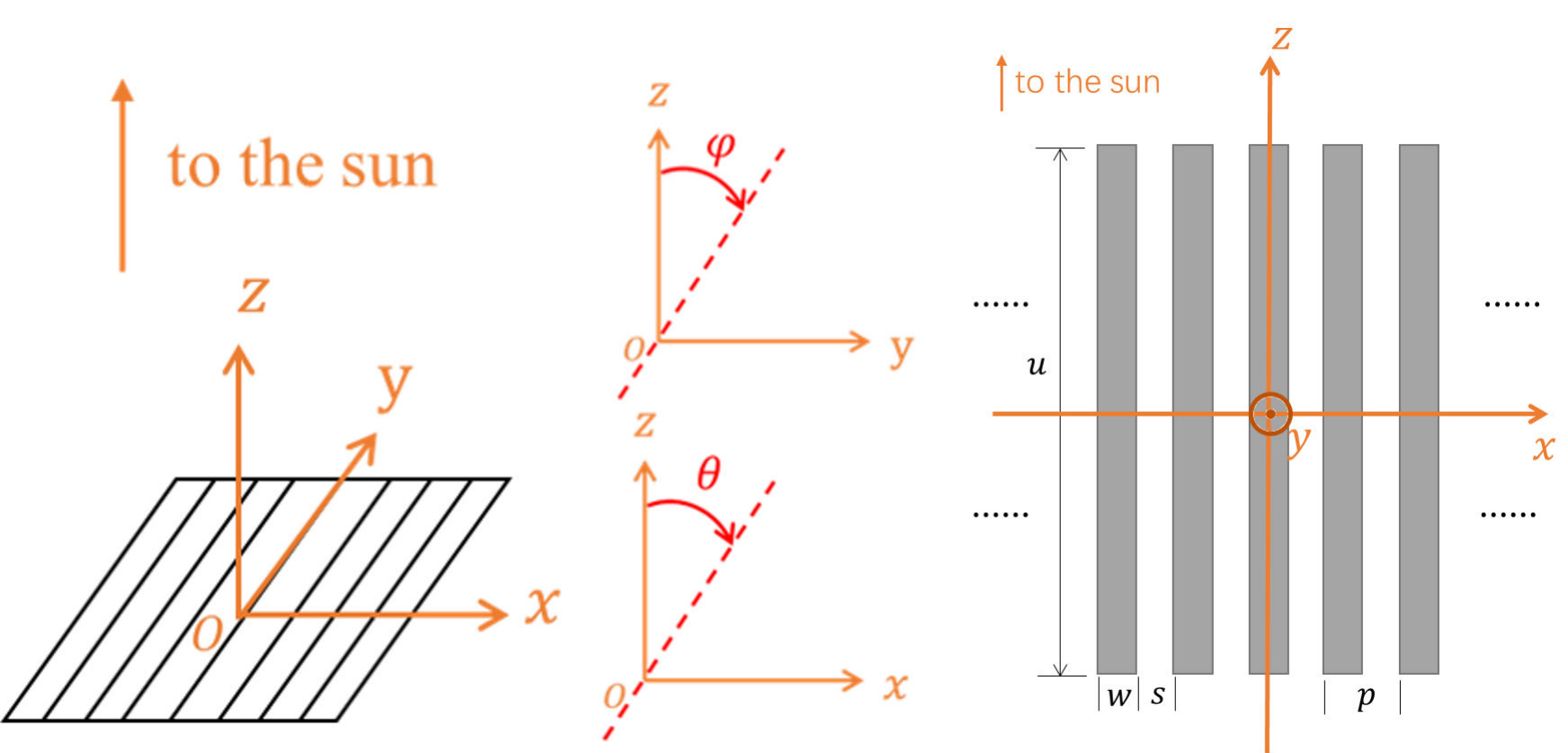}
  \caption{ The definition of coordinate system and incident angles, notice that the point $\left(0,0\right)$ is defined in the middle of a slat. 
  The arrows mark the positive direction of them.  \label{fig:coor}}
\end{figure}

We need to work out the transmittance function for a single grid in this part.
In the center of the grid, a coordinate system is built and the definition of the incidence angle of photons is specified (see Fig.\ref{fig:coor}).
Notice that $y$ and $\varphi$ can be ignored when analyzing the grid because of the translational symmetry in its $y$ direction.
Thus, the transmittance function can be written as $T(x,\theta)$, 
where $x$ is the photon's incidence point and $\theta$ is its incidence angle.

Here, this formula is applied to photons with an energy $E$ to work out $T(x,\theta)$:
\begin{equation} 
  T(x,\theta)=\exp\left[-\frac{t(x,\theta)}{\lambda_E}\right],
\end{equation}
where $t(x,\theta)$ is the length of tungsten that photons with incidence position $x$ and incidence angle $\theta$ pass through. 
And $\lambda_E$ is the photons' radiation length in tungsten at energy $E$. 
Based on the geometric relationship, when $\theta\ll 1$, $t(x,\theta)$ can be written as:
\begin{equation}
  t(x,\theta) =
  \begin{cases} 
    \frac{w+u\lv\theta\rv-2(x\bmod p)}{2\lv\theta\rv}
    & \frac{\lv u\lv\theta\rv-w \rv}{2} < x\bmod p < \frac{p-\lv u\lv\theta\rv-s \rv}{2}\\
    \frac{u\lv\theta\rv-s+\lv u\lv\theta\rv-s \rv}{2\lv\theta\rv}
    & \frac{p-\lv u\lv\theta\rv-s \rv}{2} < x\bmod p < \frac{p+\lv u\lv\theta\rv-s \rv}{2}\\
    \frac{w+u\lv\theta\rv-2p+2(x\bmod p)}{2\lv\theta\rv}
    & \frac{p+\lv u\lv\theta\rv-s \rv}{2} < x\bmod p < \frac{2p-\lv u\lv\theta\rv-w \rv}{2}\\
    \frac{w+u\lv\theta\rv-\lv u\lv\theta\rv-w \rv}{2\lv\theta\rv}
    & else
  \end{cases}.
\end{equation}
Here, $w$(wire) is the width of the tungsten slat, $s$ stands for the width of the slit,
$p=s+w$ is the pitch of the grid, and $u$ for the thickness of the grid. 
This equation only works when $\lv\theta\rv\le\frac{p}{u}$, which is met in most cases for a sub-collimator. 
So we have $T(x,\theta)$ as:
\begin{equation}
  T(x,\theta) =
  \begin{cases} 
    \exp\left[-{\frac{w+u\lv\theta\rv-2(x\bmod p)}{2\lv\theta\rv{\lambda_E}}}\right]
    & \frac{\lv u\lv\theta\rv-w \rv}{2} < x\bmod p < \frac{p-\lv u\lv\theta\rv-s \rv}{2}\\
    \exp\left[-{\frac{u\lv\theta\rv-s+\lv u\lv\theta\rv-s \rv}{2\lv\theta\rv{\lambda_E}}}\right]
    & \frac{p-\lv u\lv\theta\rv-s \rv}{2} < x\bmod p < \frac{p+\lv u\lv\theta\rv-s \rv}{2}\\
    \exp\left[-{\frac{w+u\lv\theta\rv-2p+2(x\bmod p)}{2\lv\theta\rv{\lambda_E}}}\right]
    & \frac{p+\lv u\lv\theta\rv-s \rv}{2} < x\bmod p < \frac{2p-\lv u\lv\theta\rv-w \rv}{2}\\
    exp\left[-{\frac{w+u\lv\theta\rv-\lv u\lv\theta\rv-w \rv}{2\lv\theta\rv{\lambda_E}}}\right]
    & else
  \end{cases}.
\end{equation}

Taking into account that the form of FT is favorable for integration works,
for $T(x,\theta)$ is a periodic and even function,
a Fourier cosine expansion can be performed to $T(x,\theta)$:
\begin{equation}
  \begin{aligned}
    g_0(\theta)=\frac{1}{p}\int_{0}^{p} T(x,\theta)\, {\rm d}x=&\frac{\left[\lv u\lv\theta\rv-w \rv-2\lv\theta\rv\lambda_E\right]}{p}
    \cdot{\exp\left[-{\frac{w+u\lv\theta\rv-\lv u\lv\theta\rv-w \rv}{2\lv\theta\rv{\lambda_E}}}\right]}\\
    &+\frac{\left[\lv u\lv\theta\rv-s \rv+2\lv\theta\rv\lambda_E\right]}{p}
    \cdot{\exp\left[-{\frac{u\lv\theta\rv-s+\lv u\lv\theta\rv-s \rv}
    {2\lv\theta\rv{\lambda_E}}}\right]},
  \end{aligned}
\end{equation}
\begin{equation}
  \begin{aligned}
    g_n(\theta)=&\frac{2}{p}\int_{0}^{p} T(x,\theta)\cos{\frac{2n\pi x}{p}}\, {\rm d}x=\frac{2p}{n\pi\sqrt{p^2+4n^2\pi^2\theta^2{\lambda_E}^2}}\\
    &\left\{{\sin{\left[\frac{n\pi\lv u\lv\theta\rv-w \rv}{p}-\tan^{-1}{\frac{2n\pi\lv\theta\rv{\lambda_E}}{p}}\right]}
    {\exp\left[-\frac{w+u\lv\theta\rv-\lv u\lv\theta\rv-w \rv}{2\lv\theta\rv{\lambda_E}}\right]}}\right.\\
    &\quad\left.-{\sin{\left[{{\frac{n\pi(p-\lv u\lv\theta\rv-s \rv)}{p}}}-\tan^{-1}{\frac{2n\pi\lv\theta\rv{\lambda_E}}{p}}\right]}
    {\exp\left[-{\frac{u\lv\theta\rv-s+\lv u\lv\theta\rv-s \rv}{2\lv\theta\rv{\lambda_E}}}\right]}}\right\} \\
  \end{aligned}
\end{equation}
So, $T(x,\theta)$ can be written as:
\begin{equation}
  T(x,\theta)=g_0(\theta)+\sum_{n=1}^\infty g_n(\theta)\cos{\frac{2n\pi x}{p}}.
\end{equation}
Here, $g_0(\theta)$ is the direct-current (DC) component of  $T(x,\theta)$, 
and $g_n(\theta)\cos{\frac{2n\pi x}{p}}$ is the alternating-current (AC) component. 
They are all the functions of $\theta$.

The modulation function of two layers of grids is then discussed, 
here a sub-collimator in which its front and rear grid share the same form of modulation function is considered.
According to Hurford's article \citep{hurford2003rhessi}, 
it can be calculated as an integration of the transmittance function $T(x,\theta)$ of the front grid 
and $T\left(x+L\theta+\frac{\theta_0}{2\pi}p,\theta\right)$ of the rear gird, if no deformation exists.
Here, $L$ is the distance between the front and rear grid,
while $\theta_0$ is the collimator phase of this sub-collimator.
In the case of periodic grids, it can be solved in one pitch: 
\begin{equation}
  P(\theta)=\frac{1}{p}\int_{0}^{p} T\left(x,\theta\right)T\left(x+L\theta+\frac{\theta_0}{2\pi}p,\theta\right)\, {\rm d}x,
\end{equation} 
By applying the integral properties of cosine functions to it, we can get: 
\begin{equation}
  P(\theta)={g_0(\theta)}^2+\sum_{n=1}^\infty\frac{{g_n(\theta)}^2}{2}\cos{\left(2n\pi\frac Lp\theta+n \theta_0\right)}.
\end{equation}
Here, ${g_0(\theta)}^2$ is the DC component of the modulation function.
Also, $\cos{\left(2n\pi\frac Lp\theta+n \theta_0\right)}$ acts as the modulation part of the AC component,
while $\frac{{g_n(\theta)}^2}{2}$ is the modulus.

\begin{figure}[ht!]
  \centering
  \includegraphics[width=0.98\textwidth]{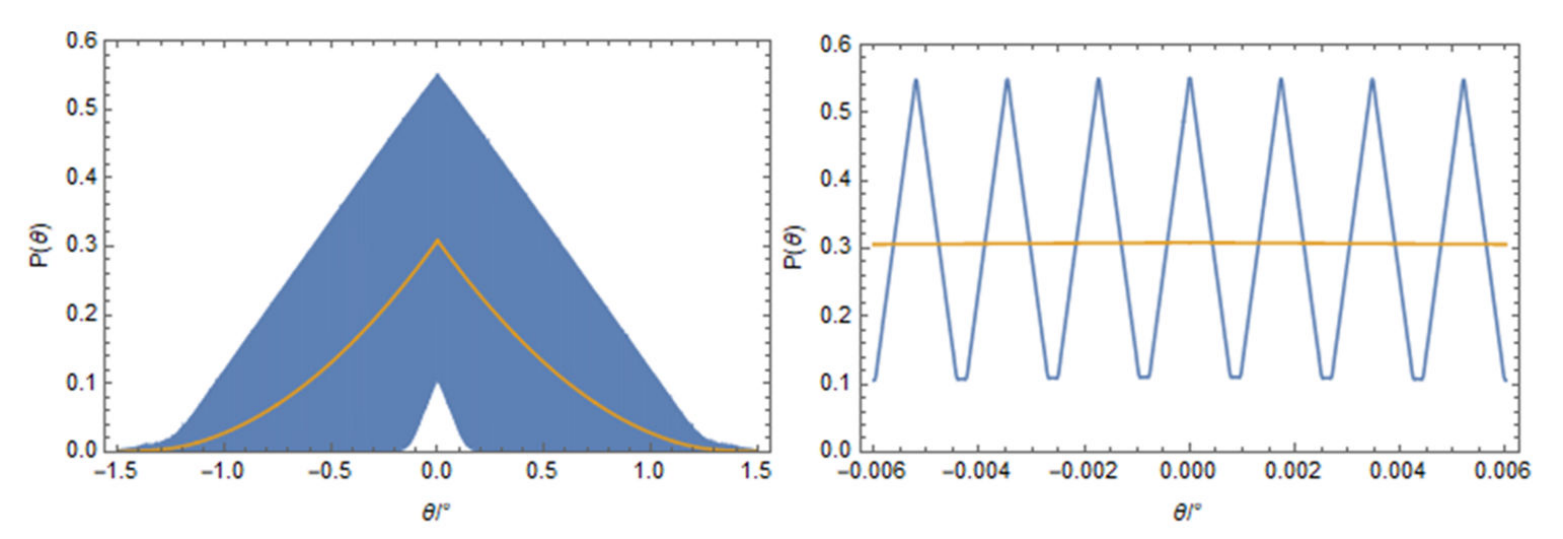}
  \caption{ The modulation function of a 36\um-pitch sub-collimator with 
  $s$=20\um, $u$=1000\um, $L$=1190mm, $\theta_0$=0, and a series summation of $n$=20. 
  The function from $-1.5^\circ$ to $1.5^\circ$ is shown on the left with the blue line, 
  while the orange line represents its DC component. 
  The right shows the detail of the center of the left curve.  \label{fig:36modul}}
\end{figure}

The modulation function of a 36\um-pitch sub-collimator with $s$=20\um, $u$=1000\um, $L$=1190mm, $\theta_0$=0,
and a series summation of $n$=20 is shown in Fig.\ref{fig:36modul}. 
It can be obtained that the DC component drops as the angle becomes larger, 
and the shape of the function changes near $0.3^{\circ}$ and $1.3^{\circ}$.
And it also shows that a summation of only 20 series can effectively handle the unequal of slit and slat.

\subsection{Degrees of Freedom (DOFs): Rigid Body} \label{subsec:rigidbody}

The modulation function of an ideal sub-collimator has been discussed above.
However, it won't be that perfect since manufacturing errors and launching deformation in position and angle may occur.
In this part, we consider the grid as a rigid body, calculate its movements in six DOFs to investigate their influence on modulation function,
and then try to simplify it under several conditions.

\subsubsection{Rotation matrix of a single layer of grid} \label{subsubsec:rotmatrix}

\begin{figure}[ht!]
  \centering
  \includegraphics[width=0.9\textwidth]{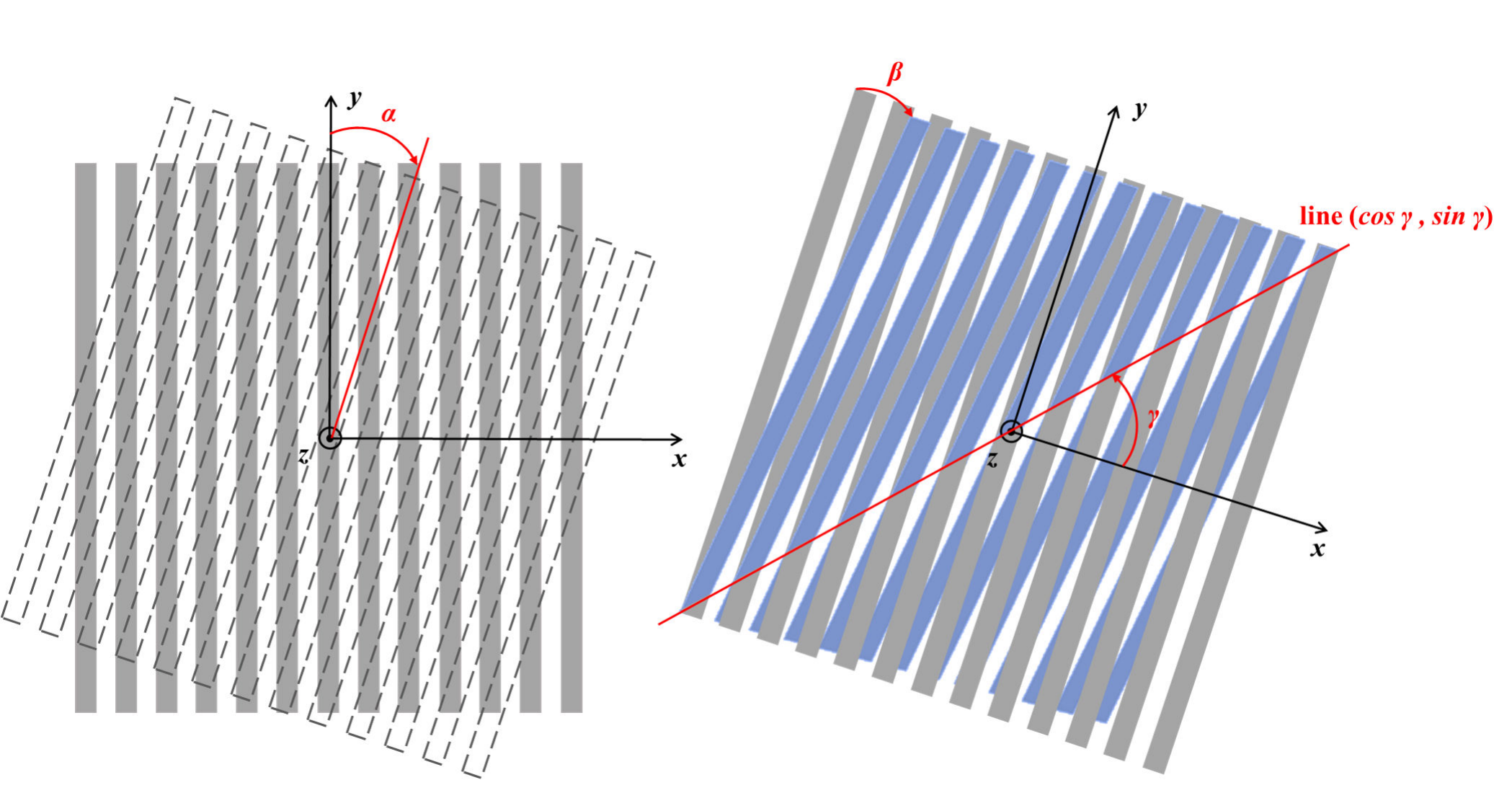}
  \caption{ The definition of rotates described by rotation matrices $A$(left) and $B$(right).}
  Here, $\alpha$, $\beta$ and $\gamma$ can describe all the rotations in space.  \label{fig:matrix}
\end{figure}

A rotation matrix is defined for the grid to help with the analysis:
First, the grid rotates clockwise by $\alpha$ degrees with its $z$-axis (matrix $A$).
Then it flips $\beta$ degrees clockwise by a line on the grid's plane (matrix $B$), 
the line has a counterclockwise angle $\gamma$ from the grid's $x$-axis (Fig.\ref{fig:matrix}).

The rotation matrix of the first rotating is matrix $A$:
\begin{equation}
  A=\begin{pmatrix}
    c_\alpha & -s_\alpha &  0\\
    s_\alpha & c_\alpha & 0\\
    0 & 0 & 1\\
  \end{pmatrix}.
\end{equation}
Here we use $c_\theta$ to represent $\cos\theta$ and $s_\theta$ for $\sin\theta$.
And the flip's rotation matrix is matrix B:
\begin{equation}
  B=\begin{pmatrix}
    c_\gamma^2+c_\beta s_\gamma^2 & (1-c_\beta)c_\gamma s_\gamma & s_\beta s_\gamma\\
    (1-c_\beta)c_\gamma s_\gamma & c_\beta c_\gamma^2+s_\gamma^2 & -s_\beta c_\gamma\\
    -s_\beta s_\gamma & s_\beta c_\gamma & c_\beta\\
  \end{pmatrix}.
\end{equation}
So here, matrix S is the rotation matrix of the whole process: 
\begin{equation}
  S=B\cdot A=\begin{pmatrix}
    c_{\alpha-\gamma}c_\gamma-c_\beta s_{\alpha-\gamma}s_\gamma & 
    -s_{\alpha-\gamma}c_\gamma-c_\beta c_{\alpha-\gamma}s_\gamma & s_\beta s_\gamma\\
    c_{\alpha-\gamma}s_\gamma+c_\beta s_{\alpha-\gamma}c_\gamma &
    -s_{\alpha-\gamma}s_\gamma+c_\beta c_{\alpha-\gamma}c_\gamma & -s_\beta c_\gamma\\
    s_\beta s_{\alpha-\gamma} & s_\beta c_{\alpha-\gamma} & c_\beta\\
  \end{pmatrix}.
\end{equation}
Compared with the Euler matrix $E$ under $(z,y,z^\prime)$ system of Euler angles $(a,b,c)$: 
\begin{equation}
  E=\begin{pmatrix}
    c_a c_b c_c-s_a s_c & -s_a c_b c_c-c_a s_c & s_b c_c\\
    s_a c_c+c_a c_b s_c & c_a c_c - s_a c_b s_c & s_b s_c\\
    -c_a s_b & s_a s_b & c_b\\
  \end{pmatrix},
\end{equation}
the relationship between $(\alpha,\beta,\gamma)$ and $(a,b,c)$ can be found as:
\begin{equation}
  \begin{cases}
    \alpha=a+c\\
    \beta=b\\
    \gamma=c+\frac{\pi}{2}\\
  \end{cases}.
\end{equation}
Therefore, it's clear that the matrix $S$ provides a complete description of the 3-dimension rotation.

\subsubsection{The projection of incident photon to the detection plane} \label{subsubsec:projection}

\begin{figure}[ht!]
  \centering
  \includegraphics[width=0.75\textwidth]{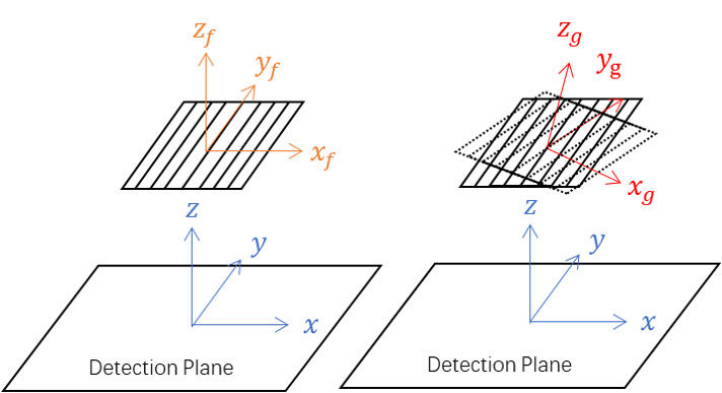}
  \caption{ The three coordinate systems are defined : the ``detection plane system" is fixed on the detector; 
  the ``system f" only moves with the grid and keeps parallel with the ``detection plane system", 
  while the ``system g" moves and also rotates with the grid.  \label{fig:coor2}}
\end{figure}

We have already shown that incident photons with position $x$ and angle $\theta$ 
have a transmittance of $T(x,\theta)$ in the previous section. 
There, $x$ and $\theta$ are defined in a coordinate system fixed to the grid, 
which we refer to as ``coordinate system $g$". 
This system follows the movement and rotation of the grid. 
The subscript ``$g$'' will be used to describe parameters under this system,
so $T(x,\theta)$ will be written as $T(x_g,\theta_g)$. 
And the ``coordinate system $f$," which moves with the grid but does not rotate with it, 
is one of the coordinate systems defined. 
Parameters in it will be described using the subscript ``$f$".
Another is the ``detection plane system", which is the coordinate system associated with
the detection plane and in which we will finally give out the results of modulation functions.
The definitions of these three systems are shown in Fig.\ref{fig:coor2}.

Starting from here, due to the effect of rotation, 
we'll describe the incident photon using four parameters $(x,y,\theta,\varphi)$ for calculation.
Here, $(x,y)$ describes the position where the photon hits the $XOY$ plane of the coordinate system, 
and a vector $(-\tan\theta,-\tan\varphi,-1)$ defines the coming position of the photon. 
And a vector $(-\theta,-\varphi,-1)$ is used to describe the photon in HXI 
because $\theta$ and $\varphi$ are $\ll$ 1.
On the $XOY$ plane of the $g$-system, it has:
\begin{equation}
  T(x_g,y_g,\theta_g,\varphi_g)=T(x_g,\theta_g).
\end{equation}

Since we are calculating $T(x,y,\theta,\varphi)$ under the detection plane system, 
the first step is to determine $T(x_f,y_f,\theta_f,\varphi_f)$. 
Here, from the system $f$ to $g$, it has:
\begin{equation}
  \begin{pmatrix}
    x_g^\prime\\y_g^\prime\\z_g^\prime\\
  \end{pmatrix}
  =S\cdot
  \begin{pmatrix}
    x_f\\y_f\\z_f\\
  \end{pmatrix}.
\end{equation}
Only the plane $XOY$ where $z_f=0$ is concerned:
\begin{equation}
  \begin{pmatrix}
    x_g^\prime\\y_g^\prime\\z_g^\prime\\
  \end{pmatrix}
  =S\cdot
  \begin{pmatrix}
    x_f\\y_f\\0\\
  \end{pmatrix}.
\end{equation}
To fit the form of $T(x_g,y_g,\theta_g,\varphi_g)$ in the $g$-system, the point 
$(x_g^\prime,y_g^\prime,z_g^\prime)$ needs to be projected to $XOY$ plane alone incident photon's vector $(-\theta,-\varphi,-1)$, so it has:
\begin{equation}
  \begin{pmatrix}
    x_g\\y_g\\0\\
  \end{pmatrix}=
  \begin{pmatrix}
    1 & 0 & -\theta_g (\theta_f,\varphi_f)\\
    0 & 1 & -\varphi_g (\theta_f,\varphi_f)\\
    0 & 0 & 0\\
  \end{pmatrix}\cdot
  \begin{pmatrix}
    x_g^\prime\\y_g^\prime\\z_g^\prime\\
  \end{pmatrix}.
\end{equation}
In the above formula, $\theta_g$ and $\varphi_g$ are the functions of $(\theta_f,\varphi_f)$.
By calculating the direction vector of the incident photon in systems g and f, the relationship between them can be described as:
\begin{equation}
  \begin{pmatrix}
    \theta_g\\\varphi_g\\1\\
  \end{pmatrix}=\frac
  {S\cdot\begin{pmatrix}
    \theta_f\\\varphi_f\\1\\
  \end{pmatrix}}
  {S_3\cdot\begin{pmatrix}
    \theta_f\\\varphi_f\\1\\
  \end{pmatrix}}.
\end{equation}
Here, $S_3$ is the $3^{rd}$ row of $S$: 
$S_3=(s_\beta s_{\alpha-\gamma},s_\beta c_{\alpha-\gamma},c_\beta)$. Finally, we get:
\begin{equation}
  \begin{pmatrix}
    x_g\\y_g\\0\\
  \end{pmatrix}=
  \begin{pmatrix}
    1 & 0 & -\theta_g (\theta_f,\varphi_f)\\
    0 & 1 & -\varphi_g (\theta_f,\varphi_f)\\
    0 & 0 & 0\\
  \end{pmatrix}\cdot
  \left[S\cdot\begin{pmatrix}
    x_f\\y_f\\0\\
  \end{pmatrix}\right].
\end{equation}
Then, assuming that the translation offset of the grid is $(x_0,y_0,z_0)$, 
and the distance of the grid to the detection plane before the translation is $L_g$, so it has:
\begin{equation}
  \begin{pmatrix}
    x_f\\y_f\\\theta_f\\\varphi_f\\
  \end{pmatrix}=\begin{pmatrix}
    x-x_0+L_g\theta+z_0\theta\\y-y_0+L_g\varphi+z_0\varphi\\\theta\\\varphi\\
  \end{pmatrix}
\end{equation}
After simplifying them, we can get:
\begin{equation}
  \begin{aligned}x_g=&\frac{(x-x_0+L_g\theta+z_0\theta)(c_{\alpha-\gamma}c_\beta c_\gamma-s_{\alpha-\gamma}s_\gamma+s_\beta c_\gamma \varphi)}
    {c_\beta+s_\beta(s_{\alpha-\gamma}\theta+c_{\alpha-\gamma}\varphi)}\\
    &-\frac{(y-y_0+L_g\varphi+z_0\varphi)(s_{\alpha-\gamma}c_\beta c_\gamma+c_{\alpha-\gamma}s_\gamma+s_\beta c_\gamma \theta)}
    {c_\beta+s_\beta(s_{\alpha-\gamma}\theta+c_{\alpha-\gamma}\varphi)},
  \end{aligned}
\end{equation}
\begin{equation}
  \begin{aligned}
    y_g=&\frac{(x-x_0+L_g\theta+z_0\theta)(c_{\alpha-\gamma}c_\beta c_\gamma+s_{\alpha-\gamma}c_\gamma+s_\beta s_\gamma \varphi)}
    {c_\beta+s_\beta(s_{\alpha-\gamma}\theta+c_{\alpha-\gamma}\varphi)}\\
    &-\frac{(y-y_0+L_g\varphi+z_0\varphi)(s_{\alpha-\gamma}c_\beta c_\gamma-c_{\alpha-\gamma}c_\gamma+s_\beta s_\gamma \theta)}
    {c_\beta+s_\beta(s_{\alpha-\gamma}\theta+c_{\alpha-\gamma}\varphi)},
  \end{aligned}
\end{equation}
\begin{equation}
  \begin{aligned}
    \theta_g=\frac{\theta(c_{\alpha-\gamma}c_\gamma-s_{\alpha-\gamma}c_\beta s_\gamma)
    +\varphi(-s_{\alpha-\gamma}c_\gamma-c_{\alpha-\gamma}c_\beta s_\gamma)+s_\beta s_\gamma}
    {c_\beta+s_\beta(s_{\alpha-\gamma}\theta+c_{\alpha-\gamma}\varphi)},
  \end{aligned}
\end{equation}
\begin{equation}
  \begin{aligned}
    \varphi_g=\frac{\theta(c_{\alpha-\gamma}s_\gamma+s_{\alpha-\gamma}c_\beta c_\gamma)
    +\varphi(-s_{\alpha-\gamma}s_\gamma+c_{\alpha-\gamma}c_\beta c_\gamma)-s_\beta c_\gamma}
    {c_\beta+s_\beta(s_{\alpha-\gamma}\theta+c_{\alpha-\gamma}\varphi)}.
  \end{aligned}
\end{equation}
They are the transformation from the detector's coordinate system to the grid's coordinate system we require.

\subsubsection{Integration of double layer of grids} \label{subsubsec:doublegrids}

Through the calculations above, $T(x,y,\theta,\varphi)$ can now be written as:
\begin{equation}
  T(x,y,\theta,\varphi)=T\left[x_g(x,y,\theta,\varphi),\theta_g(x,y,\theta,\varphi)\right]=
  \sum_{n=0}^\infty {g_n\left[\theta_g(\theta,\varphi)\right]\cos{\frac{2n\pi x_g(x,y,\theta,\varphi)}{p}}}.
\end{equation}
The term $g_0$ is inserted into the summation term to ease the statement and now the series are summed up from zero. 
To solve the modulation function of two layers of grids, we will substitute the subscript $g$ in the preceding 
equations with $t$(top) and $b$(bottom) to identify the function of the front plane from the rear plane:
\begin{equation}
  T_t=\sum_{m=0}^\infty {t_m\left[\theta_t(\theta,\varphi)\right]\cos{\frac{2m\pi x_t(x,y,\theta,\varphi)}{p}}},
  T_b=\sum_{n=0}^\infty {b_n\left[\theta_b(\theta,\varphi)\right]\cos{\frac{2n\pi x_b(x,y,\theta,\varphi)}{p}}}.
\end{equation}
The front grid has a set of parameters that are described by the subscript $t$: 
$(\alpha_t,\beta_t,\gamma_t,x_{0t},$ $y_{0t},z_{0t},L_t,s_t,w_t,p_t,u_t)$, whereas the back grid's parameters 
are described by the subscript $b$: $(\alpha_b,\beta_b,\gamma_b,x_{0b},y_{0b},z_{0b},L_b,s_b,w_b,p_b,u_b)$.

Assuming that the detection area is $D$, the modulation function can be calculated as: 
\begin{equation}
  \begin{aligned}
    P(\theta,\varphi)&=\frac{\iint_{D} \,T_t(x,y,\theta,\varphi) T_b(x,y,\theta,\varphi) 
    \mathrm{d}x\,\mathrm{d}y}{\iint_{D} \, \mathrm{d}x\,\mathrm{d}y}\\
    &=\frac{1}{\iint_{D} \, \mathrm{d}x\,\mathrm{d}y}\sum_{m,n=0}^\infty{t_m(\theta_t)b_n(\theta_b)
    {\iint_{D} \,\cos{\frac{2m\pi x_t}{p_t}}\cos{\frac{2n\pi x_b}{p_b}} \mathrm{d}x\,\mathrm{d}y}}
  \end{aligned}
\end{equation}
For the calculation of the integration ${\iint_{D} \,\cos{\frac{2m\pi x_t}{p_t}}\cos{\frac{2n\pi x_b}{p_b}} \mathrm{d}x\,\mathrm{d}y}$,
we write $\frac{2m\pi x_t(x,y,\theta,\varphi)}{p_t}$ as $a_m x+b_m y+c_m$, and $\frac{2n\pi x_b(x,y,\theta,\varphi)}{p_b}$ as $a_n x+b_n y+c_n$.
And we define $a_{mn\pm}=a_m\pm a_n,b_{mn\pm}=b_m\pm b_n,c_{mn\pm}=c_m\pm c_n$.
Then the integration can be solved under several conditions.

First, to consider a rectangle detection area of length $w_a$ and width $w_b$, the result is:
\begin{equation}
  \begin{aligned}
    P(\theta,\varphi)=&\sum_{m,n=0}^\infty{P_{mn}(\theta,\varphi)}\\
    =&\sum_{m,n=0}^\infty\frac{2t_m(\theta_t)b_n(\theta_b)}{w_a w_b}\cdot
    \left\{\frac{\sin\left({w_a a_{mn+}}/{2}\right)\sin\left({w_b b_{mn+}}/{2}\right)\cos(c_{mn+})}{a_{mn+}b_{mn+}}\right.\\
    &\qquad\qquad\qquad\qquad\quad
    \left.+\frac{\sin\left({w_a a_{mn-}}/{2}\right)\sin\left({w_b b_{mn-}}/{2}\right)\cos(c_{mn-})}{a_{mn-}b_{mn-}}\right\}
  \end{aligned}
\end{equation}
This result may have singular points in several values like 
$a_{mn\pm}=0$ or $b_{mn\pm}=0$, here their values are defined as limitations:
\begin{equation}
  P_{mn}(\theta,\varphi)|_{a_{mn\pm}=0}=\lim_{a_{mn\pm} \to 0} P_{mn}(\theta,\varphi),
  P_{mn}(\theta,\varphi)|_{b_{mn\pm}=0}=\lim_{b_{mn\pm} \to 0} P_{mn}(\theta,\varphi). 
\end{equation}
Then consider a round detection area with a radius $r$:
\begin{equation}
  \begin{aligned}
    &P(\theta,\varphi)=\sum_{m,n=0}^\infty{P_{mn}(\theta,\varphi)}\\
    &=\sum_{m,n=0}^\infty
    {
      t_m(\theta_t)b_n(\theta_b)
      \left[
        \cos(c_{mn+})\frac{J_1\left(r\sqrt{a_{mn+}^2+b_{mn+}^2}\right)}{r\sqrt{a_{mn+}^2+b_{mn+}^2}}+
        \cos(c_{mn-})\frac{J_1\left(r\sqrt{a_{mn-}^2+b_{mn-}^2}\right)}{r\sqrt{a_{mn-}^2+b_{mn-}^2}}
      \right]
    }
  \end{aligned}
\end{equation}
Here, $J_1(x)$ is the Bessel-J function of the first order. 
Also, the values of singular points are defined as limitations:
\begin{equation}
  \begin{aligned}
    P_{mn}(\theta,\varphi)|_{a_{mn+}=b_{mn+}=0}=\lim_{(a_{mn+}^2+b_{mn+}^2) \to 0} P_{mn}(\theta,\varphi),\\
    P_{mn}(\theta,\varphi)|_{a_{mn-}=b_{mn-}=0}=\lim_{(a_{mn-}^2+b_{mn-}^2) \to 0} P_{mn}(\theta,\varphi).
  \end{aligned}
\end{equation}

\subsubsection{Approximation under engineering conditions} \label{subsubsec:engineering}
The rear grid and the front grid share the same parameter in HXI for a specific sub-collimator, 
that is $u_t=u_b=u$, $w_t=w_b=w$, $s_t=s_b=s$ and $p_t=p_b=p$. 
And the detection area is limited by the diagram of the rear grid: a round area with $r$=11mm. 
Then, taking engineering parameters into consideration, it can be found that the item
$\frac{J_1\left(r\sqrt{a_{mn+}^2+b_{mn+}^2}\right)}{r\sqrt{a_{mn+}^2+b_{mn+}^2}}|_{m^2+n^2\neq 0}$ and 
$\frac{J_1\left(r\sqrt{a_{mn-}^2+b_{mn-}^2}\right)}{r\sqrt{a_{mn-}^2+b_{mn-}^2}}|_{m\neq n}$ are really small (no greater than $O(10^{-4}))$. 
Only when m=n=0,$\frac{J_1\left(r\sqrt{a_{00+}^2+b_{00+}^2}\right)}{r\sqrt{a_{00+}^2+b_{00+}^2}}=\frac{1}{2}$, 
it becomes the DC component. So, it finally has:
\begin{equation}
  P(\theta,\varphi)=g_0(\theta_t)g_0(\theta_b)+\sum_{n=1}^\infty {
  g_n(\theta_t)g_n(\theta_b)\frac{J_1\left(r\sqrt{a_{nn-}^2+b_{nn-}^2}\right)}{r\sqrt{a_{nn-}^2+b_{nn-}^2}}\cos(c_{nn-})
  }.
\end{equation}

\begin{figure}[ht!]
  \centering
  \includegraphics[width=0.98\textwidth]{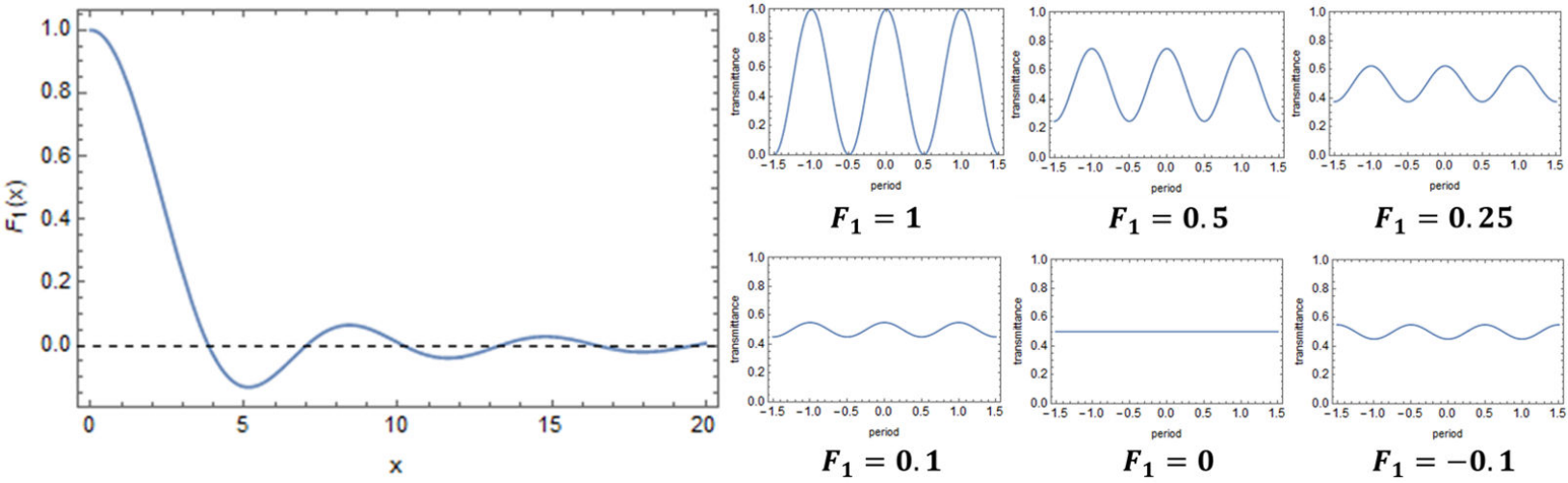}
  \caption{ The shape of the ``Modulation Factor" $F_1 (x)$ and its effects on modulation $T=\frac{1}{2}+\frac{1}{2}F_1\cos⁡\theta$. \label{fig:modulfact}}
\end{figure}

Next, we define $\alpha_h=\frac{\alpha_t+\alpha_b}{2}$, $\Delta\alpha=\alpha_t-\alpha_b$, $L=L_t-L_b$, 
$\Delta x=x_{0t}-x_{0b}$, $\Delta y=y_{0t}-y_{0b}$ and $\Delta z=z_{0t}-z_{0b}$. 
In HXI $\Delta\alpha, \beta_t, \beta_b$ are all satisfying small angle approximation ($\ll 1$), so $P(\theta,\varphi)$ can be approximated as:
\begin{equation}
  \begin{aligned}
    P(\theta,\varphi)&=g_0(\theta_t)g_0(\theta_b)\\
    &+\sum_{n=1}^\infty{
    \frac{g_n(\theta_t)g_n(\theta_b)}{2}
    \frac{J_1\left[2n\pi({r}/{p})\Delta\alpha\right]}{n\pi({r}/{p})\Delta\alpha}
    \cos{\frac{2n\pi\left[
      (\theta c_{\alpha_h}-\varphi s_{\alpha_h})(L+\Delta z)-c_{\alpha_h}\Delta x+s_{\alpha_h}\Delta y
    \right]}{p}}
    }.\end{aligned}
\end{equation}
Here, $g_0(\theta_t)g_0(\theta_b)$ is the DC component of $P(\theta,\varphi)$,
$\frac{2n\pi(-c_{\alpha_h}\Delta x+s_{\alpha_h}\Delta y)}{p}$ is its phase angle, 
and $\frac{g_n(\theta_b)g_n(\theta_t)}{2}\frac{J_1\left[2n\pi({r}/{p})\Delta\alpha\right]}{n\pi({r}/{p})\Delta\alpha}$
is the coefficient of the Fourier series. 
Here, we define $F_n(x)=\frac{2J_1(nx)}{nx}$, it always has $|F_n(x)|\leq 1$, and ${\lim_{x \to\infty} F⁡_n\left(x\right)}=0$.
So, it is called the “modulation factor”, and the series term becomes an amplitude 
$\frac{g_n(\theta_t)g_n(\theta_b)}{2}$ times the modulation factor $F_n(2\pi\frac{r}{p}\Delta\alpha)$. 
Fig.\ref{fig:modulfact} shows the shape of $F_1(x)$ with its effects on modulation: $T=\frac{1}{2}+\frac{1}{2}F_1\cos⁡\theta$.
It can be seen that, as $F(x)$ falls below zero, 
the phase of the modulation function flips upside down, contributing 180 degrees to the phase angle.

Here, it can be found that, for the same $x$, the higher the value of $n$, the faster $F_n(x)$ tends to 0.
And for the same $n$, the larger the value of $a$, the faster $F_n(ax)$ decreases to zero.
Because of these, as long as the term $F_n(ax)$ exists, the series' higher-order terms tend to converge quickly.

We can keep simplifying the function if some further assumptions are made: 
When there does not have incline angle, which is $\beta_t=\beta_b=0$, it has $\theta_b=\theta_t=\theta_g=\theta c_h-\varphi s_h$. 
Then $P(\theta,\varphi)$ can be simplified as:
\begin{equation}
  \begin{aligned}
    P(\theta,\varphi)&=g_0^2(\theta c_h-\varphi s_h)\\
    +&\sum_{n=1}^\infty{
    \frac{g_n^2(\theta c_h-\varphi s_h)}{2}
    \frac{J_1\left[2n\pi(r/p)\Delta\alpha\right]}{n\pi(r/p)\Delta\alpha}
    \cos{\frac{2n\pi\left[
        (\theta c_{\alpha_h}-\varphi s_{\alpha_h})(L+\Delta z)-c_{\alpha_h}\Delta x+s_{\alpha_h}\Delta y
      \right]}{p}
    }}.\end{aligned}
\end{equation}
Also, when there has no twist (but still can incline), $\Delta\alpha=0$, $c_{\alpha_h}=c_\alpha$, it has:
\begin{equation}
  P(\theta,\varphi)=g_0(\theta_t)g_0(\theta_b)+\sum_{n=1}^\infty{
  \frac{g_n(\theta_t)g_n(\theta_b)}{2}
  \cos{\frac{2n\pi\left[
    (\theta c_{\alpha_h}-\varphi s_{\alpha_h})(L+\Delta z)-c_{\alpha_h}\Delta x+s_{\alpha_h}\Delta y
  \right]}{p}}
}.
\end{equation}

\subsection{Behavior under Artificial X-ray Sources} \label{subsec:sources}
\begin{figure}[ht!]
  \centering
  \includegraphics[width=0.9\textwidth]{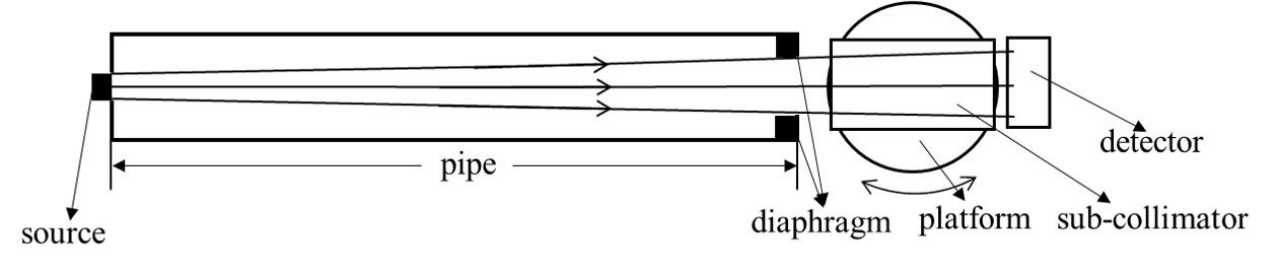}
  \caption{ This scheme shows the placement of the ground X-ray beam instruments with the collimator and detector.   \label{fig:beam}}
\end{figure}

The behavior of the sub-collimator under artificial sources will be described in this section. 
To verify this model, the HXI needs to be characterized by an X-ray beam for its modulation functions.
However, the parallel X-ray beam is quite difficult to get on the ground,
so we normally produce the X-ray beam by hitting a molybdenum target with accelerated electrons, 
then limiting the beam's divergence via pipe and diaphragms.
So here, we will discuss an X-ray source with a launch area of a radius $r_s$, 
a pipe length $l$, and a diaphragm of a radius $R$ at the pipe's end. 
As shown in Fig.\ref{fig:beam}, the collimator is put on a rotatable platform 
so that the incident angle of the X-ray beam can be changed step by step during the test.

\begin{figure}[ht!]
  \centering
  \includegraphics[width=0.55\textwidth]{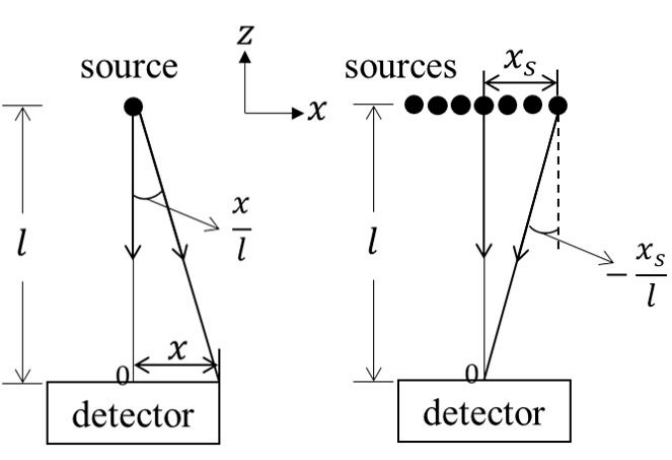}
  \caption{ The definition of position and incident angle of non-parallel light and extended source.
   Here we assume that the small angle approximation is met.   \label{fig:beam2}}
\end{figure}

Consider the situation where there is only non-parallel light. 
The incident angles $(\theta,\varphi)$ for parallel light do not change with the incident position $(x,y)$.
However, with non-parallel light, assuming an incidence angle of $(\theta,\varphi)$ at the center $(0,0)$, as shown in Fig.\ref{fig:beam2},
the incident angle of photons hitting point $(x,y)$ will be $(\theta_x,\varphi_y)=(\theta+\frac{x}{l},\varphi+\frac{y}{l})$. 
Then we can have the integration:
\begin{equation}
  P(\theta,\varphi)=P_0^\prime+\sum_{n=1}^\infty {
  \frac{1}{2}\frac{1}{\pi R^2}
  {
    \iint_{x^2+y^2\leq R^2} \,g_n^2\left[\theta_g(\theta_x,\varphi_x)\right]
    \cos(a_{nn-}^\prime x)\cos(b_{nn-}^\prime y)\cos(c_{nn-}^\prime) \mathrm{d}x\,\mathrm{d}y
  }
}.
\end{equation}
In this integration, the high-frequency terms are ignored and superscript ${}^\prime$ are used
to indicate the terms that contain $(\theta_x,\varphi_y)$. 
Notice that $P_0^\prime$ here is an integration of:
\begin{equation}
  P_0^\prime=\frac{1}{\pi R^2}{
  \iint_{x^2+y^2\leq R^2} \,g_0^2\left[\theta_g(\theta_x,\varphi_x)\right] \mathrm{d}x\,\mathrm{d}y
}.
\end{equation}
Apply the mean value theorem to $P(\theta,\varphi)$, and consider that in the experiment angles $(\theta_x,\varphi_y)$ are quite small, 
$P(\theta,\varphi)$ can be estimated as:
\begin{equation}
  \begin{aligned}
    P(\theta,\varphi)\approx&g_0^2(\theta c_h-\varphi s_h)+\sum_{n=1}^\infty
    \frac{g_n^2(\theta c_h-\varphi s_h)}{2}
    \frac{J_1\left[2n\pi\frac{R}{p}\sqrt{{(L+\Delta z)^2}/{l^2}+(\Delta\alpha)^2}\right]}
    {n\pi\frac{R}{p}\sqrt{{(L+\Delta z)^2}/{l^2}+(\Delta\alpha)^2}}\\
    &\qquad\qquad\qquad\qquad\quad\cdot\cos{\frac{2n\pi\left[
      (\theta c_{\alpha_h}-\varphi s_{\alpha_h})(L+\Delta z)-c_{\alpha_h}\Delta x+s_{\alpha_h}\Delta y
    \right]}{p}}
  \end{aligned}
\end{equation}

Then we start to consider the extended source. 
A source with uniform intensity distribution can be treated as an integration of point sources.
Assuming that the center of a launch area of radius $r_s$ has an incident angle of $(\theta,\varphi)$. 
As a result, the incident angle of the source with a position of $(x_s,y_s)$ is $(\theta-\frac{x_s}{l},\varphi-\frac{y_s}{l})$,
this is also shown in Fig.\ref{fig:beam2}.
We can approximate $P_s(\theta,\varphi)$ while $r_s$ isn't large ($\frac{r_s}{l}\ll 1$):
\begin{equation}
  \begin{aligned}
    P(\theta,\varphi)&\approx \frac{1}{\pi r_s^2}\iint_{x_s^2+y_s^2\leq r_s^2} \,
    {P\left(\theta-\frac{x_s}{l},\varphi-\frac{y_s}{l}\right)}\mathrm{d}x_s\,\mathrm{d}y_s\\
    &\approx g_0^2(\theta c_h-\varphi s_h)+\sum_{n=1}^\infty 
    \frac{g_n^2(\theta c_h-\varphi s_h)}{2}
    \frac{J_1\left[2n\pi\frac{R}{p}\sqrt{{(L+\Delta z)^2}/{l^2}+(\Delta\alpha)^2}\right]}
    {n\pi\frac{R}{p}\sqrt{{(L+\Delta z)^2}/{l^2}+(\Delta\alpha)^2}}\\
    &\qquad\qquad\cdot\frac{J_1\left[2n\pi r_s(L+\Delta z)/(p\,l)\right]}
    {n\pi r_s(L+\Delta z)/(p\,l)}\cos{\frac{2n\pi\left[
      (\theta c_{\alpha_h}-\varphi s_{\alpha_h})(L+\Delta z)-c_{\alpha_h}\Delta x+s_{\alpha_h}\Delta y
    \right]}{p}}.
  \end{aligned}
\end{equation}
Because $\Delta z$ is so small in the beaming experiment, we can let $L+\Delta z\approx L$. 
Also, we can have $\sqrt{\frac{(L+\Delta z)^2}{l^2}+(\Delta\alpha)^2}\approx\frac{L}{l}$ when $\Delta\alpha\ll\frac{L}{l}$. 
Thus, the modulation function can be written in a more straightforward form:
\begin{equation}
  P(\theta,\varphi)\approx g_0^2(\theta_g)+\sum_{n=1}^\infty \frac{g_n^2(\theta_g)}{2}
  F_n\left(2\pi\frac{L}{p\,l}R\right)F_n\left(2\pi\frac{L}{p\,l}r_s\right)
  \cos\left(2n\pi\frac{L}{p}\theta_g+n\theta_0\right).
\end{equation}
Here,
\begin{equation}
  \begin{aligned}
    &F_n\left(2\pi\frac{L}{p\,l}R\right)=\frac{J_1\left[2n\pi LR/(p\,l)\right]}{n\pi LR/(p\,l)},
    F_n\left(2\pi\frac{L}{p\,l}r_s\right)=\frac{J_1\left[2n\pi Lr_s/(p\,l)\right]}{n\pi Lr_s/(p\,l)},\\
    &\theta_0=\frac{2\pi}{p}\left(-c_{\alpha_h}\Delta x+s_{\alpha_h}\Delta y\right).
  \end{aligned}
\end{equation}
The name ``modulation factor" is used to describe $F_n\left(2\pi\frac{L}{p\,l}R\right)$ and $F_n\left(2\pi\frac{L}{p\,l}r_s\right)$.
In many cases, $2\pi\frac{L}{pl}$ is so large that $F_n$ will quickly decline to zero even when $n$ is 2. 
So, we may utilize this approximation to analyze them:
\begin{equation}
  P(\theta,\varphi)\approx g_0^2(\theta_g)+\frac{g_1^2(\theta_g)}{2}
  F_1\left(2\pi\frac{L}{p\,l}R\right)F_1\left(2\pi\frac{L}{p\,l}r_s\right)
  \cos\left(2\pi\frac{L}{p}\theta_g+\theta_0\right).
\end{equation}
Here, $F_1\left(2\pi\frac{L}{p\,l}R\right)$ is referred to as the ``factor of the diagram", 
while $F_1\left(2\pi\frac{L}{p\,l}r_s\right)$ as the ``factor of the source".
Thus, by adding the modulation factor, it allows non-parallel light to be treated similarly to parallel light. 
The effect of non-parallel light is a composite of the modulation function under parallel light with the modulation factor. 
So, the modulation function under parallel light can be obtained by simply measuring the 
modulation function under non-parallel light and deducting the effect of the modulation factor.

\section{Simulations and Experiments for collimator of HXI} \label{sec:testment}
The structure of HXI with its collimator is described in this part, 
followed by simulations using Geant4, a Monte Carlo simulation software for high-energy physics.
Then we present the X-ray beam experiments with the results. 
To prove the mathematical model, all of these works are compared to the calculation results.

\subsection{Structure of HXI} \label{subsec:structure}

\begin{figure}[ht!]
  \centering
  \includegraphics[width=0.75\textwidth]{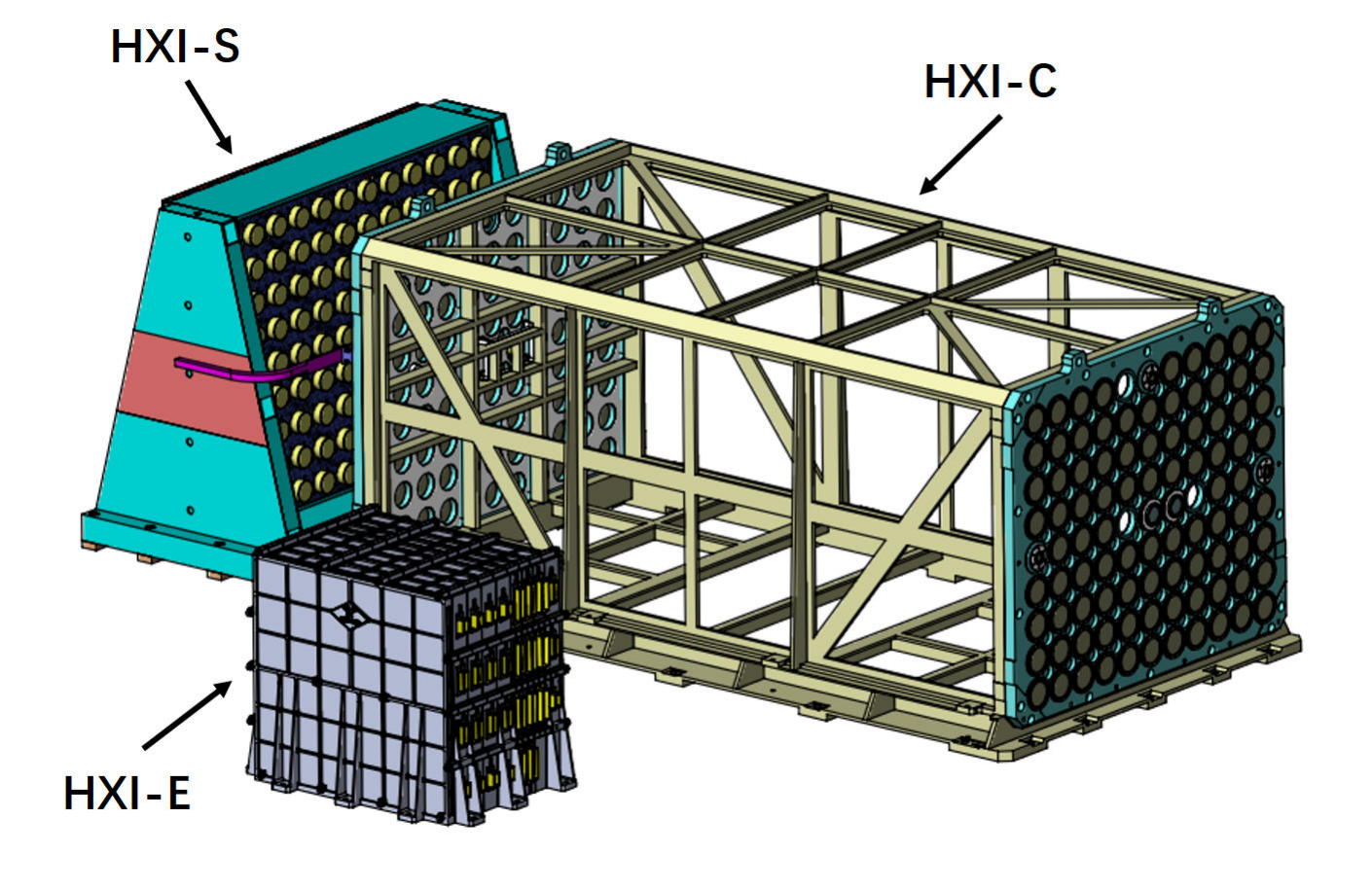}
  \caption{  The structure of HXI onboard ASO-S.  \label{fig:hxi}}
\end{figure}

Fig.\ref{fig:hxi} depicts the structure of HXI. It consists of three major parts: 
a collimator (HXI-C) performs physical Fourier-transformation for incident photons,
a spectrometer (HXI-S) detects the spectrum of transmitted photons,
and an electronics control box (HXI-E) provides power while simples data processing in orbit \citep{zhang2019hard}.

\begin{figure}[ht!]
  \centering
  \includegraphics[width=0.75\textwidth]{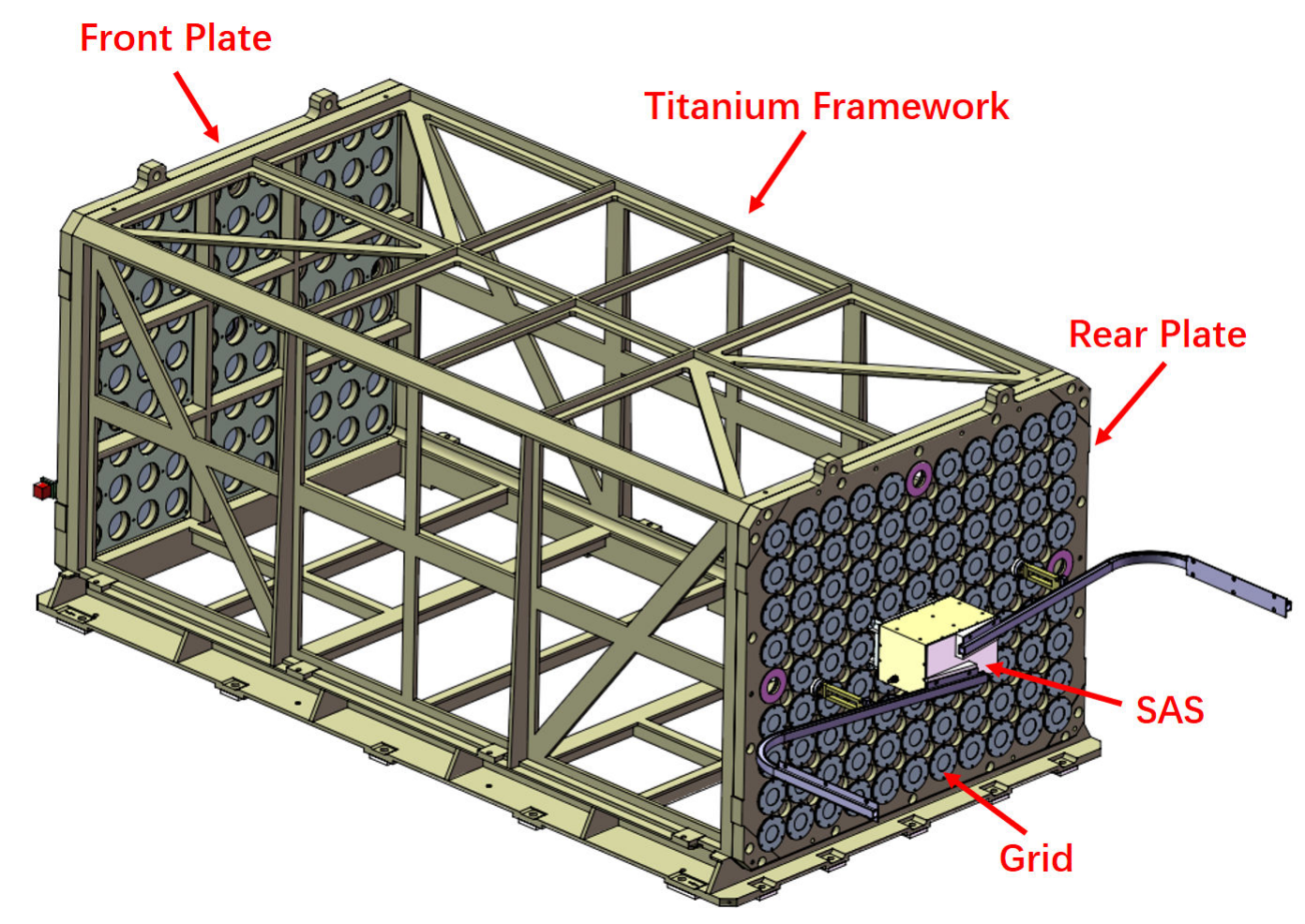}
  \caption{ The structure of HXI-C.  \label{fig:hxic}}
\end{figure}

The HXI-C (Fig.\ref{fig:hxic}), which modulates X-ray photons, is made up of two plates mounted on a titanium (Ti) framework.
The plates have an array of 11×9, however only 91 sub-collimators are set for modulation. 
The rest of the 8 holes are reserved for the measurement of the collimator's pointing and particle background. 
The Solar Aspect System (SAS), which is mounted on the rear plate, 
assists in locating the Sun's center and monitoring distortion between the front and rear plates\citep{chen2021design}.

\begin{figure}[ht!]
  \centering
  \includegraphics[width=0.75\textwidth]{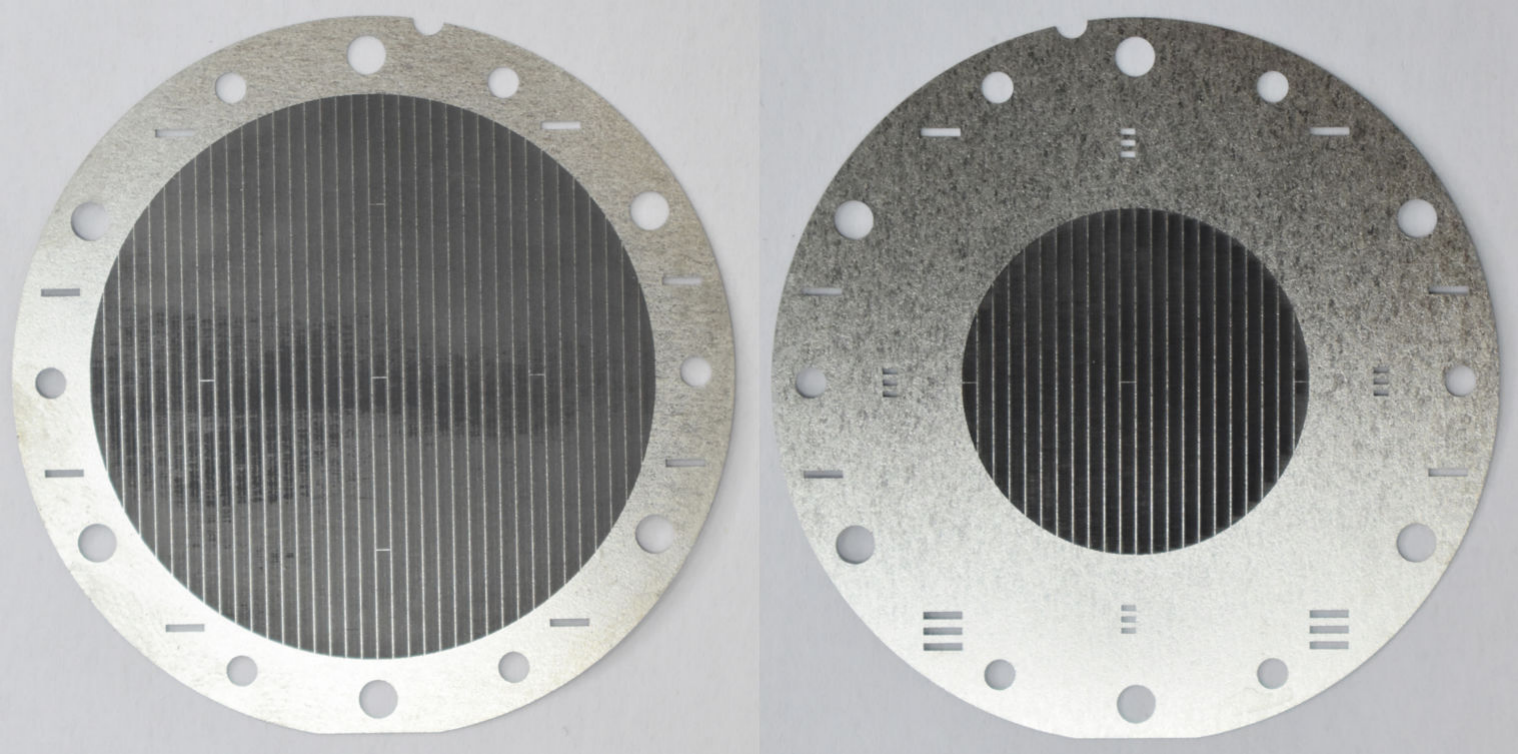}
  \caption{ The slit array of grids and its outer rings. 
  A front grid with a 36mm diameter is shown on the left, with a rear grid with a 22mm diameter on the right.  \label{fig:grids}}
\end{figure}

Each sub-collimator is comprised of two grids that are arranged in the same position as the front and rear plates, 
with a distance of 1190 mm between them. Different sub-collimators' grids have different pitches and rotation angles,
allowing them to function as different u-v parameters in frequency space. 
The grids, which are made of tungsten foils, are held together by tungsten rings (see Fig.\ref{fig:grids}).
The rings set a limit on the effective area of grids, which is 36mm in diameter for front-plate grids and 22mm for rear-plate grids.
And the detectors assembled in HXI-S have a round detection area with a diameter of 25mm.

\subsection{Simulation of modulation function in Geant4} \label{subsec:simulation}

In this section, we'll go over several Geant4 simulation results.
We want to show that the mathematical model works effectively in solving near-ideal situations.

\begin{figure}[ht!]
  \centering
  \includegraphics[width=0.98\textwidth]{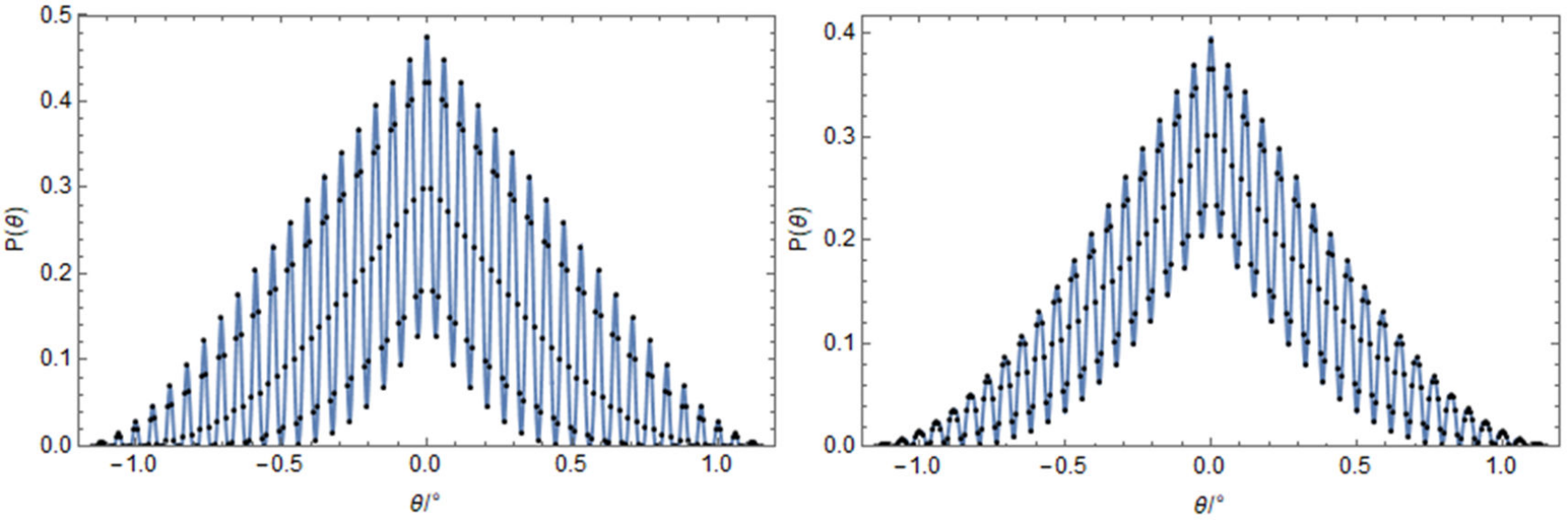}
  \caption{ The modulation functions of grids with $p$=36\um, $s$=20\um, $L$=35mm, 
  $l$=26m, $d=2r_s$=0.4mm, $D=2R$=10mm(left)/20mm(right), $\alpha$=$\beta$=0, $\theta_0$=0 and a series summation of n=2. 
  The black points represent simulation results, while the blue line is the calculated curve.  \label{fig:36short}}
\end{figure}

First, we look at a situation of 36\um-pitch grids with $p$=36\um, $s$=20\um, $L$=35mm, $l$=26m, $d=2r_s=$0.4mm, $\alpha$=$\beta$=0 and $\theta_0$=0. 
Here, the front grid of it is placed only 35mm away from the rear grid, allowing us to see the change of the function with angle $\theta$ more clearly.
And then we put it in the beaming tube to see how it behaves under experiments. 
It's worth noticing that the series in the modulation factor isn't small enough to exclude high-order terms. 
As a result, we still need to compute the summary of the first two orders.
We simulated it in Geant4, and the computation and simulation results are shown in Fig.\ref{fig:36short}. 
All of the simulations below were run under 30 keV energy.

\begin{figure}[ht!]
  \centering
  \includegraphics[width=0.98\textwidth]{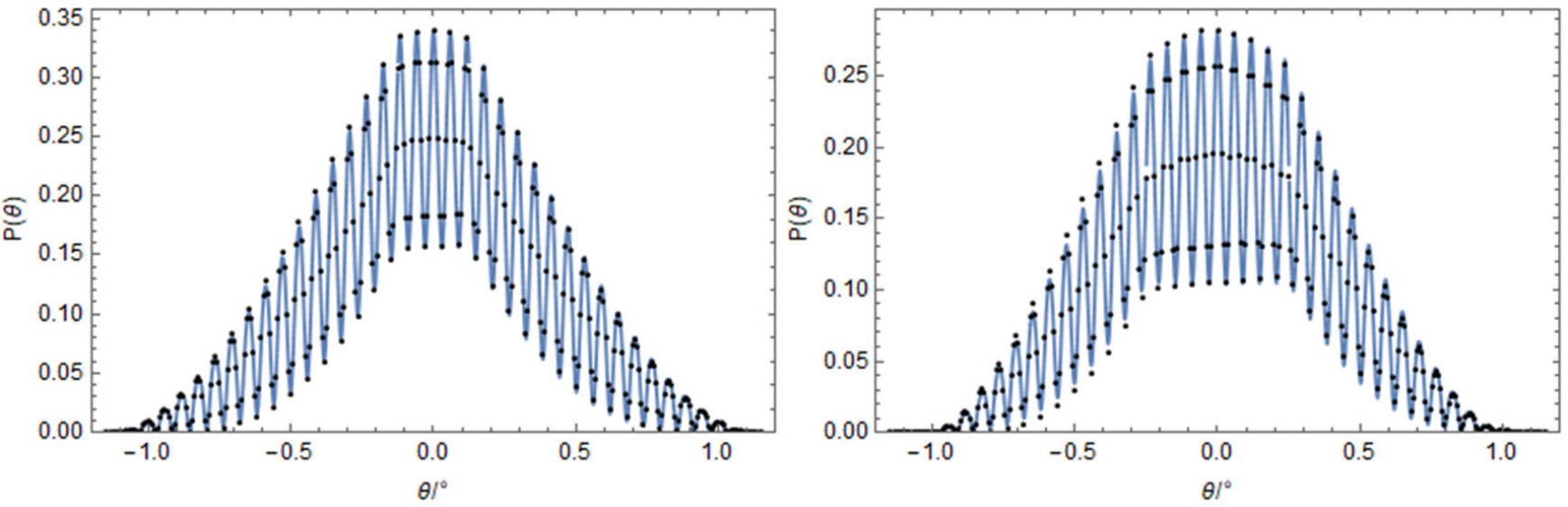}
  \caption{ The modulation functions of grids with $p$=36\um, $s$=20\um, $L$=35mm, $l$=26m, 
  $d=2r_s$=0.4mm, $D=2R$=20mm, $\alpha$=0, $\gamma=\frac{\pi}{2}$, $\beta$=$15^\prime$(left)/$30^\prime$(right). 
  The black points represent simulation results, while the blue line is the calculated curve.  \label{fig:36dip}}
\end{figure}

We can see that, in this case, the function cannot be represented as a simply cosine function, but must be treated as a summary of terms.
After that, we apply the inclination to it, as shown in Fig.\ref{fig:36dip}.
Compared to the functions without inclination in Fig.\ref{fig:36short}, it can be found that the inclination effect 
lowers the function's maximum value and brings a platform to it, and the length of the platform is the inclination angle $\beta$.

\begin{figure}[ht!]
  \centering
  \includegraphics[width=0.75\textwidth]{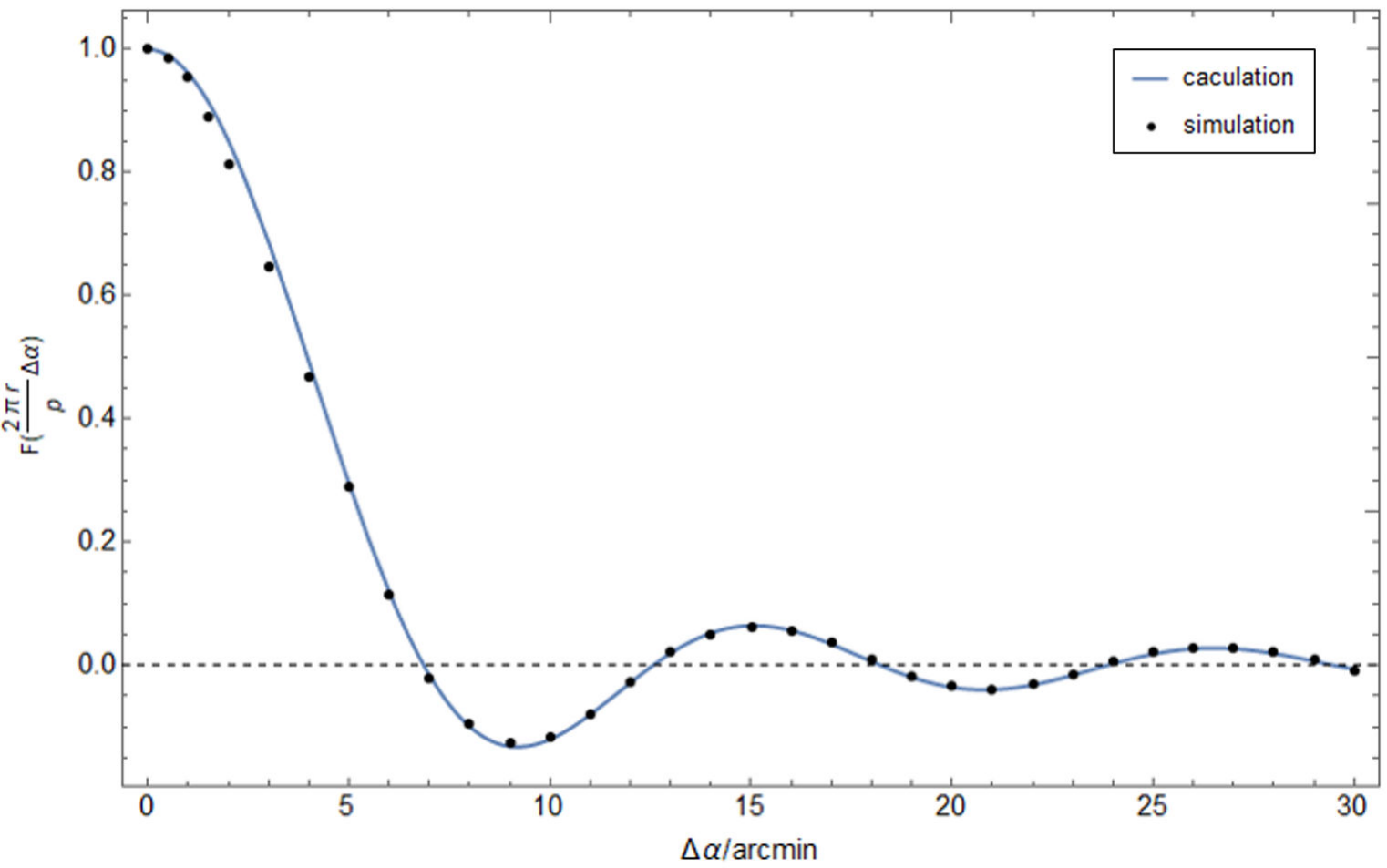}
  \caption{ $F_1\sim\Delta\alpha$ curve for 36\um-pitch sub-collimator with $r$=11mm.  \label{fig:twistcurve}}
\end{figure}

The twist angle $\alpha$ is then taken into account.
This time, we'll look at the situation in parallel light, for twist angle will be one of the most affected parameters during launch. 
We want to evaluate how big of an impact it has on the modulation factor $F_n(2\pi\frac{r}{p}\Delta\alpha)$.
Here in $F_n(2\pi\frac{r}{p}\Delta\alpha)$, those high-order terms decay quickly, so we just consider $F_1(2\pi\frac{r}{p}\Delta\alpha)$.
We can deduce from the form of $F_1(2\pi\frac{r}{p}\Delta\alpha)$ that, for the same $\Delta\alpha$, the smaller $p$ is, the larger its effect would be. 
So, in Fig.\ref{fig:twistcurve}, we analyze a 36\um-pitch sub-collimator on HXI and plot its $F_1\sim\Delta\alpha$ curve.

It can be noticed from the curve that the first zero point of the curve occurs at roughly 7 arcmins for 36\um-pitch grids. 
And $F_1$ is around 0.96 for a 1 arcmin twist. 
As a result, we can conclude that its effect on the 36\um-pitch can be ignored as long as the twist angle $\Delta\alpha$ is kept under 1 arcmin.

\begin{figure}[ht!]
  \centering
  \includegraphics[width=0.75\textwidth]{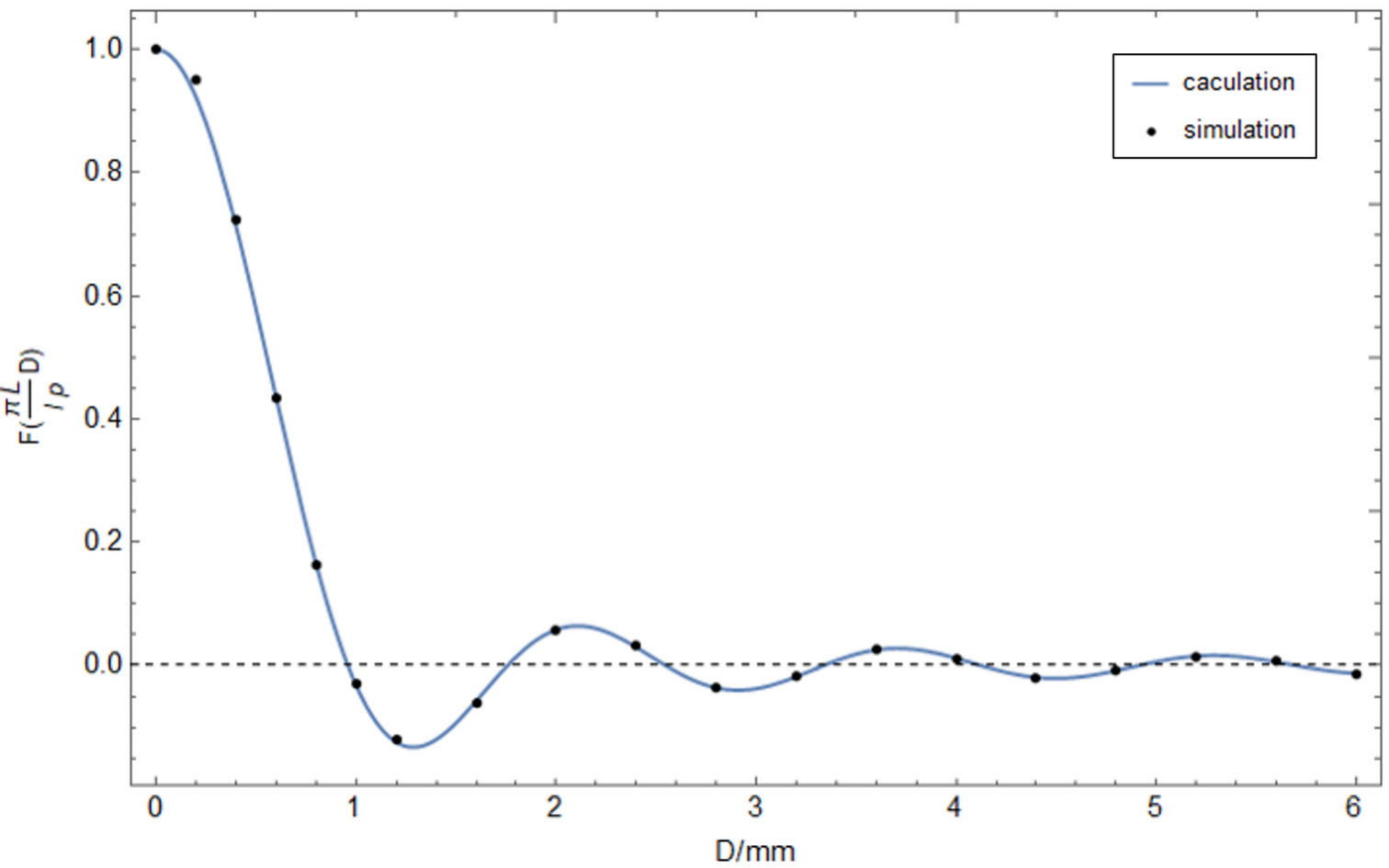}
  \caption{ Function $F_1(\pi\frac{L}{p\,l}D)\sim D$. Here, $p$=36\um, $l$=26m and $L$=1190mm.  \label{fig:diacurve}}
\end{figure}

We also want to check how the ``factor of the diagram" $F_1\left(2\pi\frac{L}{p\,l}R\right)$ 
and ``factor of the source" $F_1\left(2\pi\frac{L}{p\,l}r_s\right)$ behave under X-ray beam experiment. 
We only show the curve of $F_1\left(2\pi\frac{L}{p\,l}R\right)$ in Fig.\ref{fig:diacurve} for they share the same form.

It can be seen from the above comparison of the math model with Geant4 results that the model works well when only considering the geometric relationship. 
However, in real-world situations, a variety of environmental factors may play a role in the outcome. 
To verify this model, more evidence from the experiment is needed.

\subsection{Verification tests of X-ray Beam Experiments} \label{subsec:experiments}

\begin{figure}[ht!]
  \centering
  \includegraphics[width=0.95\textwidth]{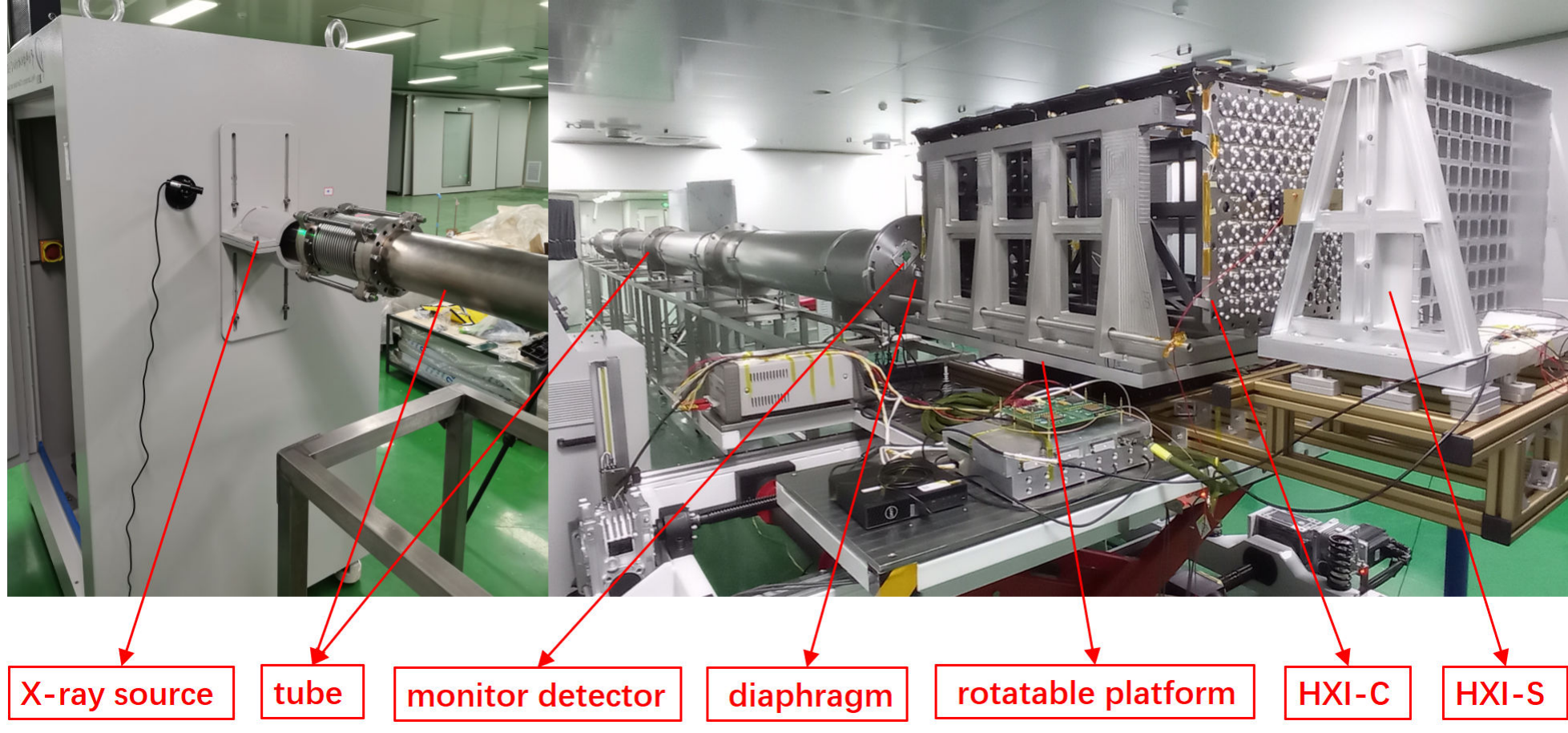}
  \caption{ The structure of the experimental devices.  \label{fig:experiment}}
\end{figure}

The verification test of HXI was performed after determining the experimental parameters by simulation works.
The experimental device's structure is depicted in Fig.\ref{fig:experiment}.
An X-ray source with an emitting area which is a disk of diameter $d=2r_s=0.4mm$ was utilized in this experiment, 
and the electronic beam's energy was turned to 30keV. 
Photons travel the length of the tube, which is $l=26m$. 
A diaphragm with $D=2R$ was placed at the end of this tube to limit the angle distribution of photons.
The size of the diaphragm could be adjusted as needed. 
A monitor was also assembled nearby to measure the intensity of the X-ray beam.
The HXI-C and HXI-S were mounted on a rotatable platform with high precision.
The HXI was rotated to change the incident angle $\theta_g$ of photons in the experiment. 

\begin{figure}[ht!]
  \centering
  \includegraphics[width=0.8\textwidth]{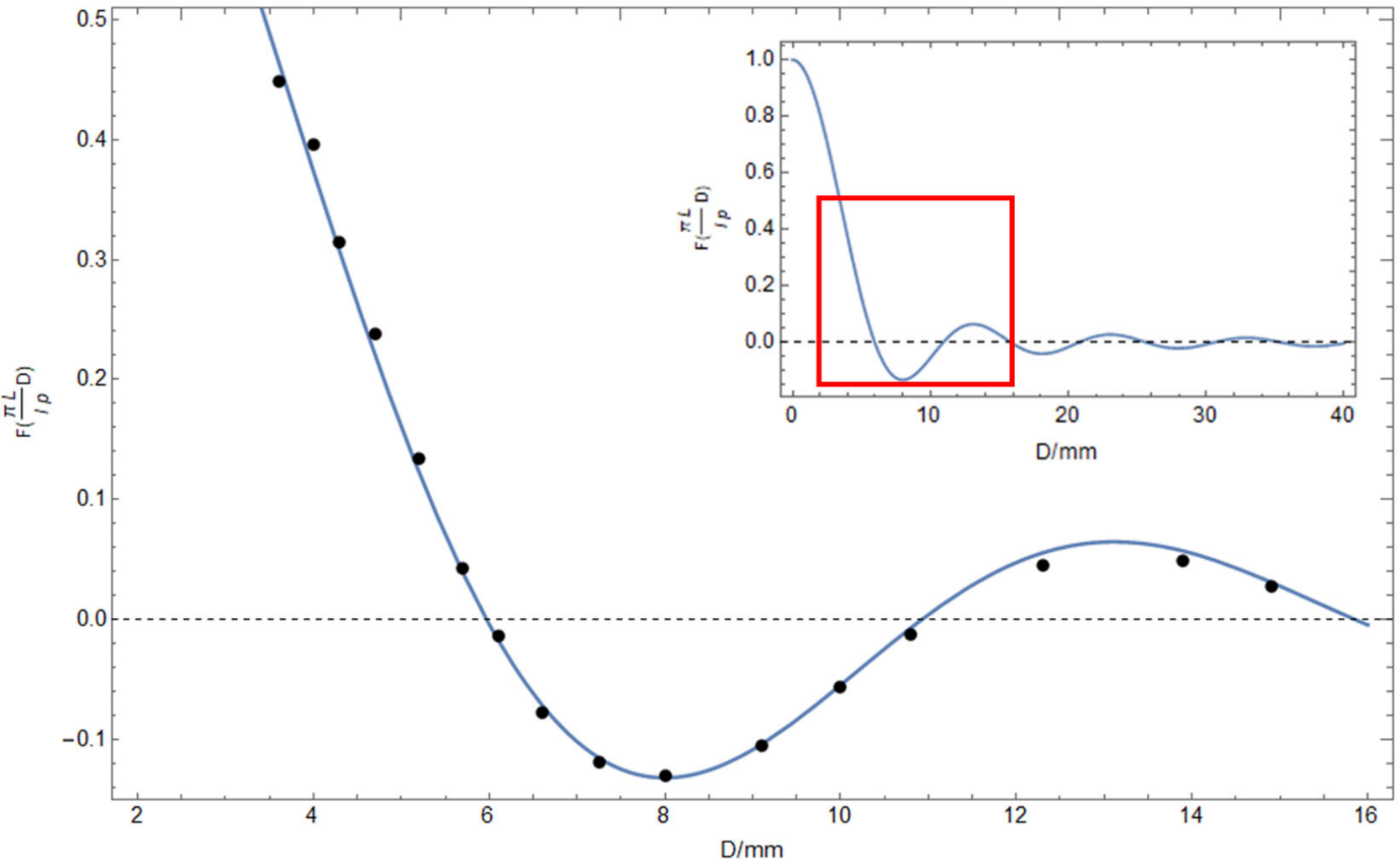}
  \caption{ The curve of $F_1(\pi\frac{L}{p\,l}D)\sim D$ for a 224\um-pitch sub-collimator with $L$=1190mm and $l$=26m.
  The value from the math model is shown by the blue line, while the black points represent the results of the experiment. 
  The top right diagram depicts the measurement region in this experiment.  \label{fig:diacurve2}}
\end{figure}

The platform could only be rotated horizontally during the experiment due to the platform's design.
As a result, $\theta_g$ becomes a function of the platform's rotation angle $\theta_p$ and rotation angle $\alpha$: $\theta_g=\theta_p\cos⁡\alpha$. 
So, the modulation function $P(\theta_p)$ on rotation angle $\theta_p$ can approximately be written as:
\begin{equation}
  P(\theta_p)\approx g_0^2(\theta_p\cos\alpha)+\frac{g_1^2(\theta_p\cos\alpha)}{2}F_1\left(\pi\frac{L}{p\,l}D\right)
  F_1\left(\pi\frac{L}{p\,l}d\right)\cos\left(2\pi\frac{L}{p}\theta_p\cos\alpha+\theta_0\right).
\end{equation}
Here, $D=2R$,$d=2r_s$. Then it came to the measure of the ``factor of the diagram".
A 224\um-pitch sub-collimator was used in this experiment. 
We first found out the function's peak area to ensure that photons can nearly incident normally. 
The diaphragm was then changed in size, 
and the platform was rotated for each diagram to measure the peak and valley of the modulation.
After a series of analyses, we were able to obtain the curve for the ``factor of the diagram", 
which is shown in Fig.\ref{fig:diacurve2}.

\begin{figure}[ht!]
  \centering
  \includegraphics[width=0.8\textwidth]{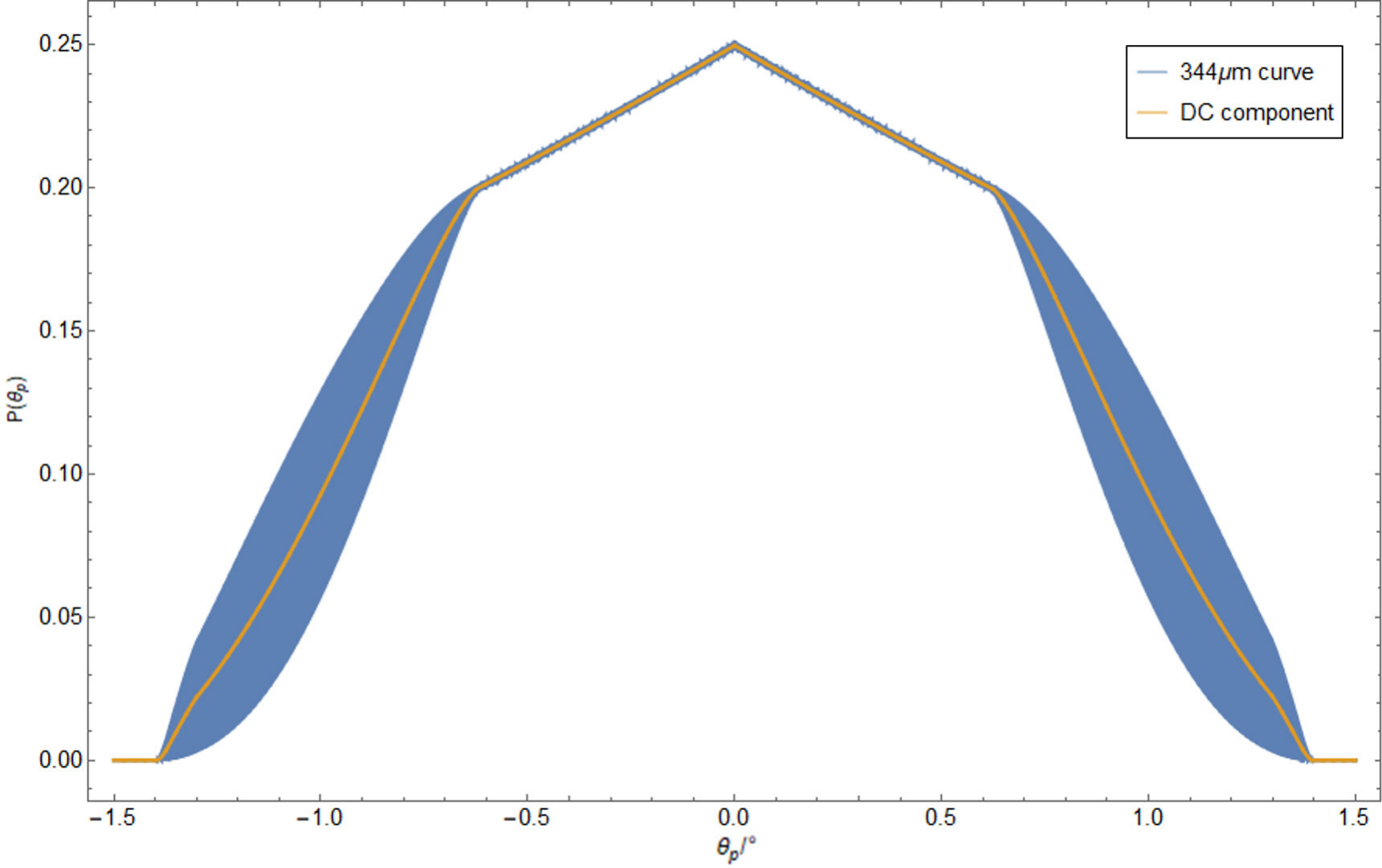}
  \caption{ The calculation curve of 344\um-pitch sub-collimator with $s$=172\um, $u$=2000\um, $D$=9.1mm, $d$=0.4mm, 
  $L$=1190mm, $l$=26m, and a rotation angle $\alpha=31^\circ$.  
  The diameter of the front grid is 36mm, 22mm for the rear grid, and 25mm for the detector.
  The curve is a contour of modulations components.  \label{fig:344curve}}
\end{figure}

After that came the measurement of the DC component for the modulation function. 
Considering the form of the function, if a suitable size for the diaphragm was chosen to make $F_1(\pi\frac{L}{p\,l}D)=0$, 
the pure DC component could be obtained from the experiment. 
But notice that, in the experiment, the incident angle was changed by rotating the platform. 
Therefore, when the angle is large, the front plane or the real plane might shadow the detector. 
Under these situations, the effect of occluding might make the detection zoom far away from a round area. 
So, the result of the integration on the round area didn't work this time, 
and the shape of the detection area even changed with the rotation of the platform.
We solved the integration by applying a changeable area with the angle, and the result can be seen in Fig.\ref{fig:344curve}.
In this figure, the calculated curve of a 344\um-pitch sub-collimator is shown with a rotation angle $\alpha=31^\circ$.
It can be seen that no occlude effect works from -0.6 to 0.6 degrees where the integration area is round and
$F_1(\pi\frac{L}{p\,l}D)\approx 0$. It means that the curve shows the DC component.
As the rotation angle becomes larger, occlude effect occurs, so the shape of the function starts to change.
The integration of the non-circular area and the amplitude of the original AC component jointly contribute to the final amplitude.

\begin{figure}[ht!]
  \centering
  \includegraphics[width=0.8\textwidth]{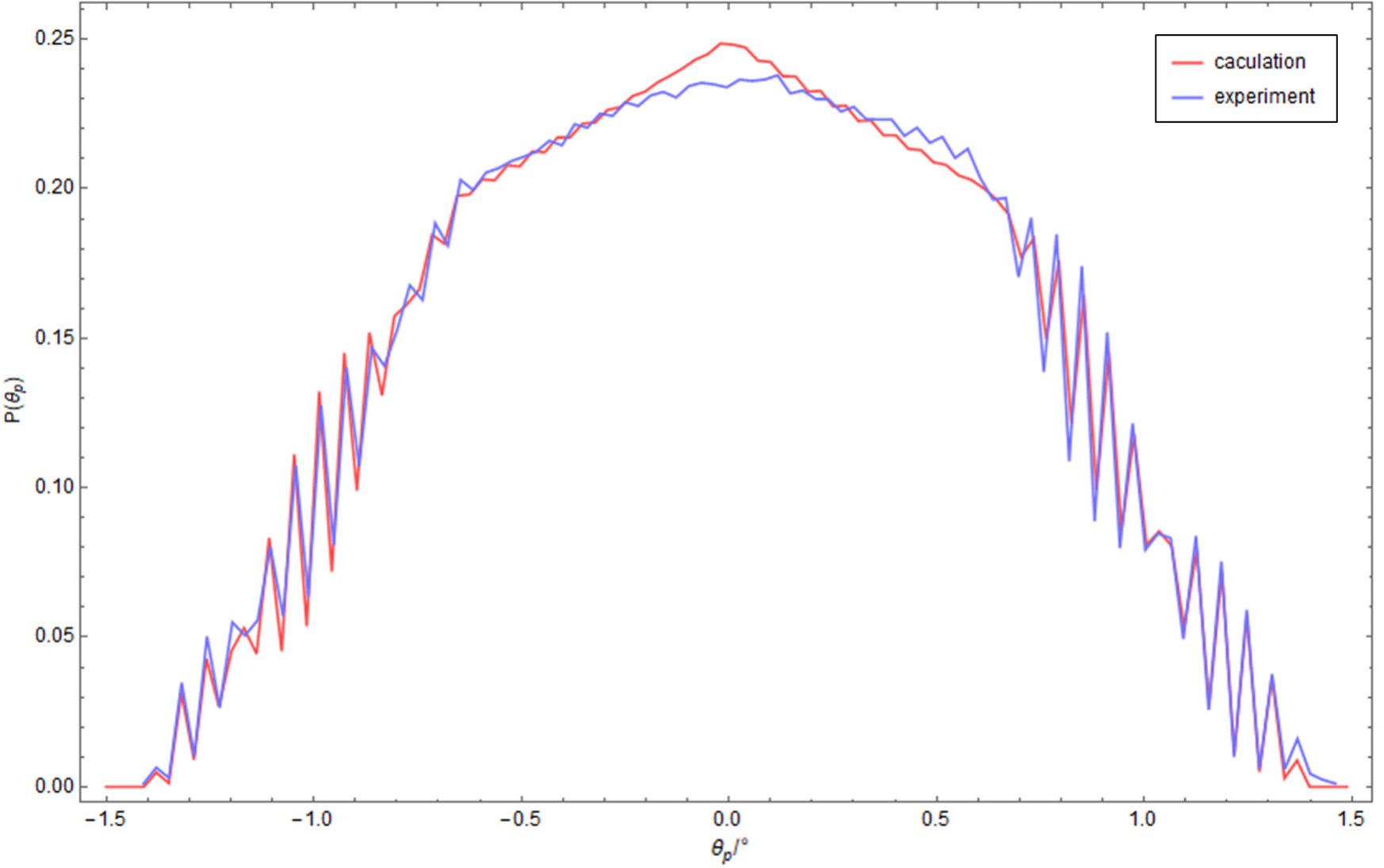}
  \caption{ The computed result of the ``DC" curve (red) compared to the experiment result (blue) with the same sampling frequency.  \label{fig:344curve2}}
\end{figure}

The modulation components in the diagram change at a really high frequency. 
Due to the time limit, it was unable to attain such a high sample frequency in the experiment. 
The measured curve will be limited by the contour of Fig.\ref{fig:344curve} if the sampling frequency is not high enough.
Fig.\ref{fig:344curve2} shows the contrast between the calculated and experimental results.
The calculation and experiment were both done with the same sampling frequency. 
These two experiments strongly show that our math model works effectively even under complex conditions.

\section{Discussion: Data process with the model} \label{subsec:discussion}

In this section, it is discussed how to consider the effects of deformation of HXI-C in the data process,
and to what extent they may affect the modulation function.
These results will present the application of the model and its superiority over other methods in calculating the pattern.

To correct the instrument deformation, the first step is to monitor the state of HXI-C with the Displacement Monitor(DM)
after it is launched into space \citep{zhang2019hard}.
It can measure the twists and shifts between the front and rear plates,
so the deviations in position and angle for each sub-collimator can be calculated. 
Then the modulation function for each bi-grid can be figured out, considering all its deformations.
As a result, the effect of deformation of HXI-C in the image reconstruction process could be weakened 
as much as possible by using the new modulation functions. 
Notice that in Section \ref{subsubsec:engineering}, an assumption of those deformation angles 
always satisfying small angle approximation has been mentioned. 
Zhang's article mentioned above has shown that during the whole mission of the satellite, 
all the deformation angles could be controlled within several arcseconds, 
and this has been verified through environmental tests.
However, as long as those angles are kept below 1 degree, they are still satisfied with the small angle approximation.
So, the simplifications in Section \ref{subsubsec:engineering} can work during the whole life of ASO-S,
since there is a huge gap between error control and assumption failure.

To show how this model works, here we suppose that the imaging area lies near the edge of the sun, 
while HXI-C points to the center of the sun.
Under this condition, the thickness of the grids is the item that affects most. 
To see its effect more clearly, we approximate the DC component of the modulation function when the energy isn't high:
${g_0(\theta)}^2\approx{\left(\frac{s-u\lv\theta\rv}{p}\right)}^2$, here we apply the assumption of $\lambda\ll u$. 
As shown with the yellow curve in Fig.\ref{fig:36modul}, the DC component drops as the incident angle becomes large.
Since the term $\frac{u}{p}\lv\theta\rv$ appears in the formula,
this effect is more pronounced for grids with larger thickness and smaller pitch.

\begin{figure}[ht!]
  \centering
  \includegraphics[width=0.98\textwidth]{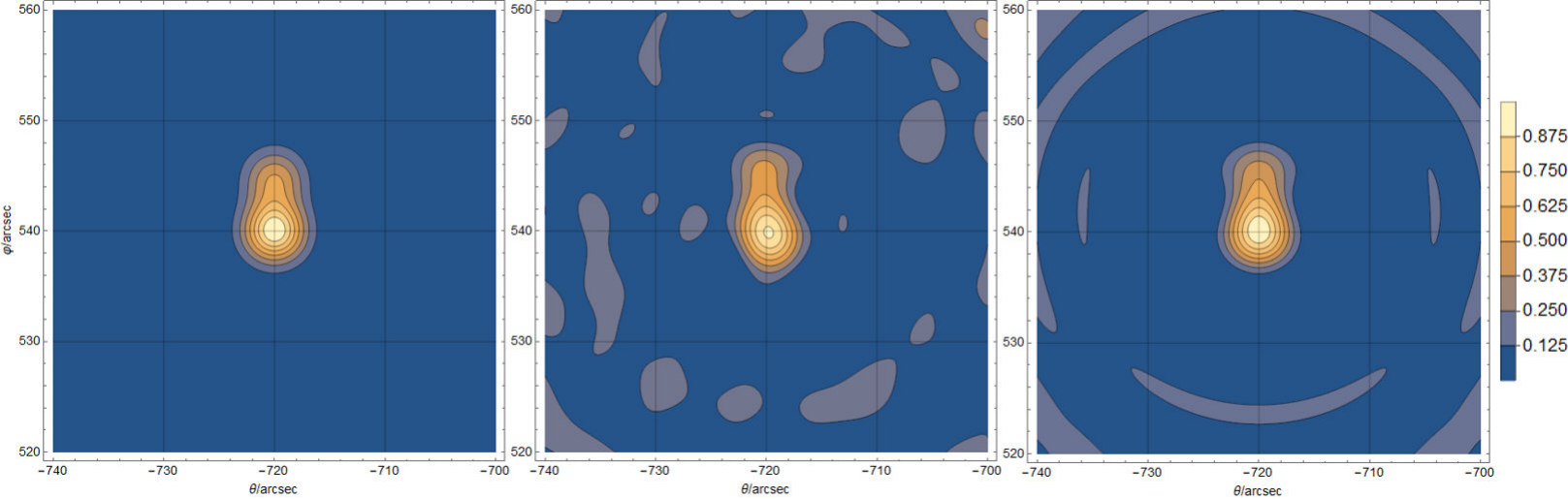}
  \caption{  The reconstructed images of the test sources,the left one gives the ideal result as a reference, the middle one is done without 
  considering the effect of thickness, while the right one is considered. \label{fig:correct}}
\end{figure}

As a test, in Fig.\ref{fig:correct}, two point sources are placed at the edge of the sun 
to see how this model helps to improve the results.
The chosen area is the upper left edge of the sun, 15 arcmins away from its center.
The two point sources - one is twice brighter than the other - are placed at a distance of 1.5x resolution, with an energy of 30 keV.
The counts for each sub-collimator are calculated, then the image is reconstructed without and with the consideration of the thickness effect. 
Notice that in the reconstruction process, for grids with different position angles, 
the incident angle $\theta$ from the same source can be different.
Therefore, each pattern needs to be calculated and corrected separately, and then recombined to form the final image.
It can be clearly seen that the shape and the brightness of the sources go wrong if the thickness of the grids isn't calculated.

\begin{figure}[ht!]
  \centering
  \includegraphics[width=0.98\textwidth]{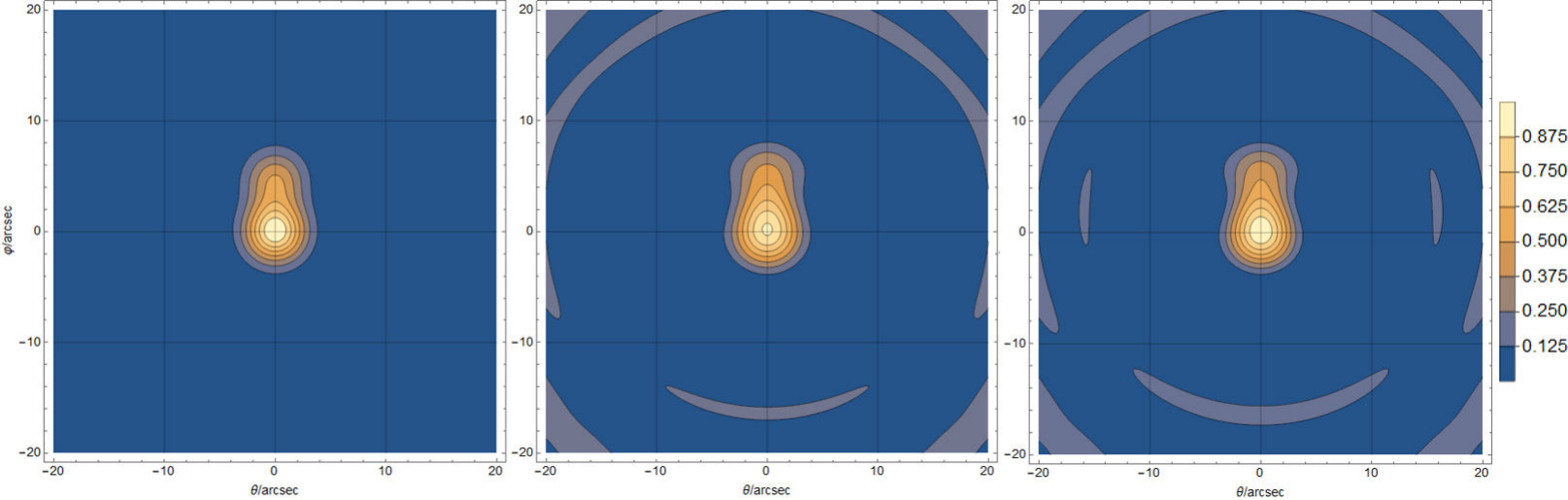}
  \caption{  The reconstructed images of the test sources, the left one gives the ideal result as a reference, the middle one is done without 
  considering the effect of twist, while the right one is considered. \label{fig:correct2}}
\end{figure}

Then it comes to the affection of twist angle $\Delta\alpha$. 
As mentioned in Section \ref{subsubsec:engineering}, this effect contributes as a modulation factor 
$F_n(2\pi\frac{r}{p}\Delta\alpha)=\frac{J_1\left[2n\pi({r}/{p})\Delta\alpha\right]}{n\pi({r}/{p})\Delta\alpha}$.
So, as a function of $\frac{r}{p}\Delta\alpha$ shown in Fig.\ref{fig:twistcurve}, for a same twist angle $\Delta\alpha$, 
its impact becomes bigger as the pitch of the grid goes smaller.
And as Fig.\ref{fig:twistcurve} shows, for the grid with a smallest 36\um-pitch in HXI, 
as long as the twist angle is smaller than 1 arcmin, its effect on the modulation can be ignored. 
If the twist angle is larger than that, the modulation function needs to be corrected.
In Fig.\ref{fig:correct2}, the same sources are used as in Fig.\ref{fig:correct}, but this time they are placed in the center of the sun.
A twist angle $\Delta\alpha=2.5\ arcmins$ is added to the instrument, 
and it can be seen that the twist makes the sources darker, but doesn't change the shape.

For the effect of inclination, as Fig.\ref{fig:36dip} shows, 
it brings a platform that has a length of the incline angle $\beta$ to the center of the modulation function.
The incident angles $\theta_t$ and $\theta_b$ are the terms that consist of the incline angle $\beta$,
and the DC component $g_0(\theta_t)g_0(\theta_b)$ is the most affected part, 
especially when $\theta_t$ and $\theta_b$ are of the same order of magnitude as $\beta$.
As a result, the effect of inclination mainly occurs when the incident angle is close to zero.
This affection mainly reduces the DC component of the modulation function, and it can be solved just like the effect of thickness.

Then, the case in which different types of effects are combined would be discussed. 
As shown in Section \ref{subsubsec:engineering}, the modulation function under engineering conditions has a form of
\begin{equation}
  \begin{aligned}
    P(\theta,\varphi)&=g_0(\theta_t)g_0(\theta_b)\\
    &+\sum_{n=1}^\infty{
    \frac{g_n(\theta_t)g_n(\theta_b)}{2}
    \frac{J_1\left[2n\pi({r}/{p})\Delta\alpha\right]}{n\pi({r}/{p})\Delta\alpha}
    \cos{\frac{2n\pi\left[
      (\theta c_{\alpha_h}-\varphi s_{\alpha_h})(L+\Delta z)-c_{\alpha_h}\Delta x+s_{\alpha_h}\Delta y
    \right]}{p}}
    }.\end{aligned}
\end{equation}

\begin{figure}[ht!]
  \centering
  \includegraphics[width=0.98\textwidth]{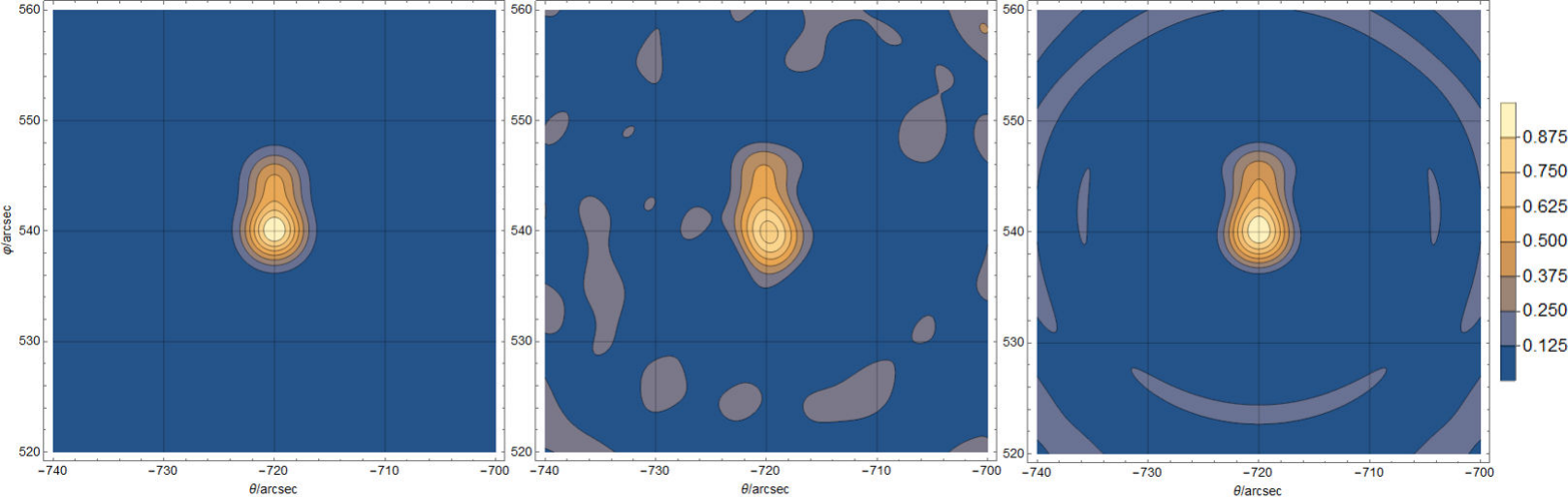}
  \caption{  The reconstructed images of the test sources, the left one gives the ideal result as a reference, the middle one is done without 
  considering the effect of thickness and twist, while the right one is considered. \label{fig:correct3}}
\end{figure}

In this expression, the terms $g_0(\theta_t)g_0(\theta_b)$ and $g_n(\theta_t)g_n(\theta_b)$ are affected both by thickness and inclination.
Fortunately, they can be distinguished more finely: $g_0(\theta)$ and $g_n(\theta_t)$ are only affected by thickness, 
while $\theta_t$ and $\theta_b$ are only affected by inclination. 
And the effect of twist, which only appears in term $\frac{J_1\left[2n\pi({r}/{p})\Delta\alpha\right]}{n\pi({r}/{p})\Delta\alpha}$,
can be easily corrected. So, both effects can be modified independently. 
To visualize this process, a case combining both effects in Fig.\ref{fig:correct} and Fig.\ref{fig:correct2} is discussed.
This time, the sources are placed at the edge of the sun, while a twist angle $\Delta\alpha=2.5\ arcmins$ is added to the instrument.
Fig.\ref{fig:correct3} shows the comparison before and after the corrections, 
it can be found that sources without corrections are darker than that in Fig.\ref{fig:correct}.
Since the effect of thickness and inclination are independent of each other, we can correct them in any order:
If the effect of thickness is corrected first, we will get the image in Fig.\ref{fig:correct2}, 
then the twist of the instrument is corrected to get the final picture.
Otherwise, the image in Fig.\ref{fig:correct} will be first gotten if the effect of twist is corrected first.
These corrections could be done at the same time and this won't make any difference to the final result.

In these cases above, although the traditional method of numerical simulation can do the same job in revising these effects,
applying the new model can be much faster. 
More importantly, this model, which can give out the mathematical form of those deformations, is much better in analyticity.

\section{Summary} \label{sec:summary}
In this paper, a new mathematical model for HXI by digging into the structure of the FT imaging telescope is given. 
This model is formed by analyzing the geometric structure of the sub-collimator, with different types of deformation taken into consideration.
The modulation functions under various conditions are calculated, 
and Monte Carlo simulations are performed to confirm the accuracy of the calculations.
A series of experiments are also designed to test this model, and the model can match the experiment results really well.
The ability of this model in solving patterns under different types of deformation is also shown.
Notice that this model is not only sufficient in analyzing the image reconstruction process of HXI,
but also quite useful with other telescopes as long as they use a similar imaging theory as HXI. 
On the other hand, we can use this model to analyze the deformation of HXI and its influences on imaging ability after launch. 
By correcting the deformations, a more accurate pattern will be provided to the imaging reconstruction process;
we believe it will be much beneficial in observing solar flares.

\normalem
\begin{acknowledgements}
  This work is supported by the Strategic Priority Research Program on Space Science, 
  Chinese Academy of Sciences (No. XDA 15320104), 
  the Scientific Instrument Developing Project of the Chinese Academy of Sciences, Grant No. YJKYYQ20200077, 
  the National Natural Science Foundation of China (Nos. 12173100 and 12022302, 11803093,11973097), 
  and the Youth Innovation Promotion Association, CAS (No. 2021317 and Y2021087). 
  The worthful work during X-ray beam experiments carried out by the HXI group is also appreciated.
\end{acknowledgements}
  
\bibliography{reference}{}
\bibliographystyle{aasjournal}

\appendix
\section{Symbol explanation} \label{sec:symbol}
\begin{center}
  \begin{table}[ht]
      \centering
      \caption{Symbols used in this paper}
      \footnotesize
      \label{tab:symbol}
      \begin{tabular}{c l c}
          \hline\hline
          Symbol(s) & Meaning(s) & Note(s) \\
          \hline
          $x$, $y$, $z$ & coordinate locations & see Fig.\ref{fig:coor} \\
          $\theta$, $\varphi$ & direction angles of incident light & see Fig.\ref{fig:coor} \\
          $u$ & the thickness of the grid & see Fig.\ref{fig:coor} \\
          $p$ & the pitch of the grid & see Fig.\ref{fig:coor} \\
          $s$ & the width of the slit in the grid & see Fig.\ref{fig:coor} \\
          $w$ & the width of the slat in the gird & see Fig.\ref{fig:coor} \\
          $\lambda_E$ & photon's radiation length in tungsten under energy E & $/$ \\
          $L$ & the distance between the front and rear grid & 1190mm for HXI-C \\
          $t$ & the length of tungsten that photons pass through & see Eq.(2) \\
          $T$ & the transmittance function for a single layer of grid & see Eq.(3) \\
          $g_n$ & the nth order coefficient of function T's Fourier expansion & see Eqs.(4-6) \\
          $P$ & pattern, the modulation function of the sub-collimator & $/$ \\
          $\alpha$ & the twist angle with the z-axis & see Fig.\ref{fig:matrix} \\
          $\beta$ & the inclination angle by line$\left(cos\gamma,sin\gamma\right)$ & see Fig.\ref{fig:matrix} \\
          $\gamma$ & the angle used to define the line by which the grid inclines & see Fig.\ref{fig:matrix} \\
          subscript$_t$ & marks the parameter of the front(top) gird & $/$ \\
          subscript$_b$ & marks the parameter of the rear(bottom) gird & $/$ \\
          subscript$_f$ & marks the parameter under “system f” & see Fig.\ref{fig:coor2} \\
          subscript$_g$ & marks the parameter under “system g” & see Fig.\ref{fig:coor2} \\
          subscript$_0$ & the basic offset for the parameter & $/$ \\
          $c_{angle}$, $s_{angle}$ & short for $cos\left(angel\right)$ and $sin\left(angel\right)$ & $/$ \\
          $a_{mn\pm}$, $b_{mn\pm}$ & intermediate quantities for simplifying expressions & see Eq.(27)\\
          $w_a$, $w_b$ & the length $w_a$ and width $w_b$ of a rectangle detection area & $/$ \\
          $r$ & the radius of a round detection area & 11mm for HXI-C \\
          $J_1\left(x\right)$ & first-order BesselJ function & $/$ \\
          $F_n\left(x\right)$ & defined as an nth-order "modulation function" & $\frac{2J_1\left(nx\right)}{nx}$ \\
          $\alpha_h$ & an average of the twist angle of the front and rear grid & $\frac{\alpha_t+\alpha_b}{2}$ \\
          $\Delta\alpha$ & the relative twist angle of the front and rear grid & $\alpha_t-\alpha_b$ \\
          $\Delta x$ & the relative offset in the x-axis between the front and rear grid & $x_{0t}-x{0b}$ \\
          $\Delta y$ & the relative offset in the y-axis between the front and rear grid & $y_{0t}-y{0b}$ \\
          $\Delta z$ & the relative offset in the z-axis between the front and rear grid & $z_{0t}-z{0b}$ \\
          $r_s$ & the radius of the emission area of the X-ray source & 0.2mm in the experiment\\
          $d$ & the diameter of the emission area of the X-ray source & $d=2r_s$ \\
          $l$ & the length of the pipe & 26m in the experiment \\
          $R$ & the radius of the diaphragm at the pipe's end & $/$ \\
          $D$ & the diameter of the diaphragm at the pipe`s end & $D=2R$ \\
          $\theta_p$ & the rotation angle of the platform in the experiment & $/$ \\
          \hline
      \end{tabular}
  \end{table}
\end{center}

\end{document}